 \documentclass[final,3p,times]{elsarticle}

\usepackage{amssymb}
\usepackage[utf8]{inputenc} 
\usepackage{subcaption}
\journal{NIMA}
\begin{document}
\begin{frontmatter}
\title{\textit{In vacuum}  diamond sensor scanner for beam halo measurements in the beam line at the KEK Accelerator Test Facility}


\author {S. Liu\fnref{label1}}
\author[label1] {F. Bogard}
\author[label1] {P. Cornebise}
\author[label2,label1] {A. Faus-Golfe}
\author[label2] {N. Fuster-Martínez}
\author[label4] {E. Griesmayer}
\author[label1] {H. Guler}
\author[label1] {V. Kubytskyi}
\author[label1] {C. Sylvia}
\author[label3] {T. Tauchi}
\author[label3] {N. Terunuma}
\author[label1] {P. Bambade}

\address[label1]{Laboratoire de l’Accélérateur Linéaire (LAL), Université Paris Sud, CNRS/IN2P3, Université Paris-Saclay, Orsay, France}
\address[label2]{Instituto de Fisica Corpuscular (CSIC-UV), Valencia, Spain}
\address[label3]{High Energy Accelerator Research Organization (KEK), Tsukuba, Japan}
\address[label4]{Atominstitut, Technische Universit\"at Wien, Wien, Austria}

\begin{abstract}
The investigation of beam halo transverse distributions is important for the understanding of beam losses and the control of backgrounds in Future Linear Colliders (FLC). A novel \textit{in vacuum} diamond sensor (DSv) scanner with four strips has been designed and developed for the investigation of the beam halo transverse distributions and also for the diagnostics of Compton recoil electrons after the interaction point (IP) of ATF2, a low energy (1.3 GeV) prototype of the final focus system for the ILC and CLIC linear collider projects. Using the DSv, a dynamic range of $\sim10^6$ has been successfully demonstrated and confirmed for the first time by simultaneous beam core ($\sim10^9$ electrons) and beam halo ($\sim10^3$ electrons) measurements at ATF2. This report presents the characterization, performance studies and tests of the diamond sensors using an $\alpha$ source as well as using the electron beams at PHIL, a low energy ($< 10$ MeV) photo-injector at LAL, and at ATF2. First beam halo measurement results using the DSv at ATF2 with different beam intensities and vacuum levels are also presented. Such measurements not only allow one to evaluate the different sources of beam halo generation but also to define the requirements for a suitable collimation system to be installed at ATF2, as well as to optimize its performance during future operation.

\end{abstract}


\end{frontmatter}


\section{Introduction}
\begin{figure*}[t]
    \centering
     \includegraphics*[width=140mm]{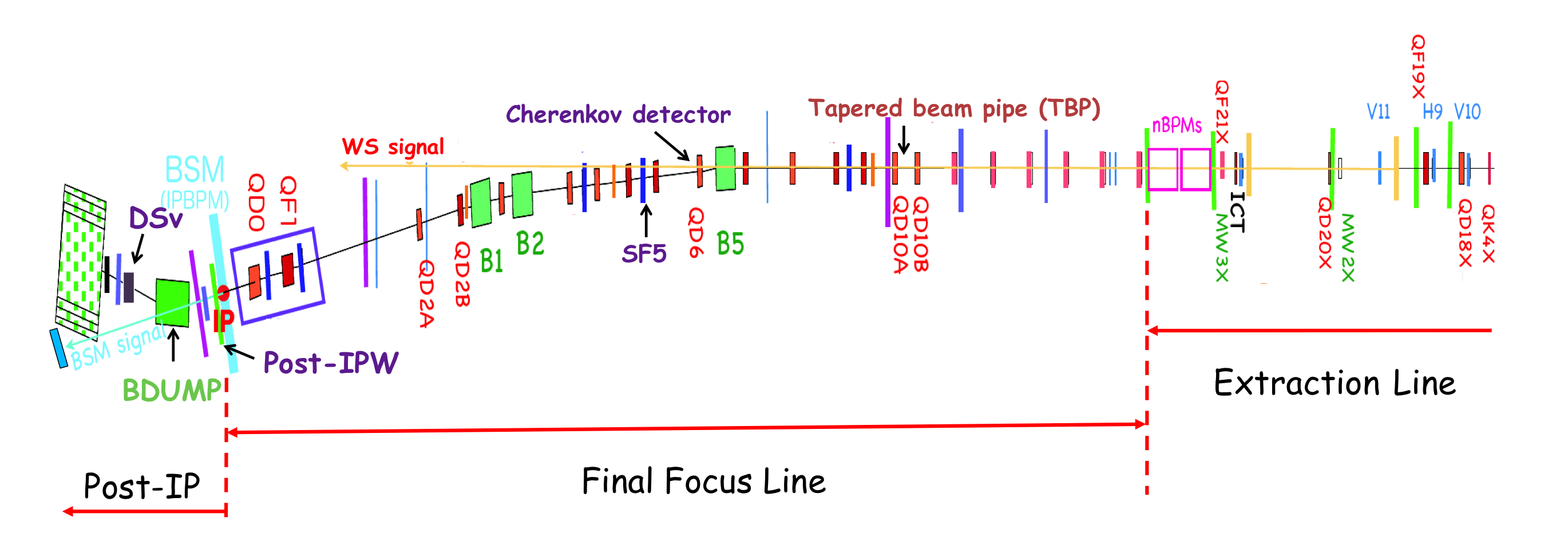}
    \caption{ATF2 layout (start from skew quadrupole QK4X): positions of the wire scanners (MW2X, MW3X and post-IPW) and the \textit{in vacuum} Diamond Sensors (DSv)  are indicated. }
    \label{fig:ATF2}
\end{figure*}

The investigation of beam halo transverse distributions is of great importance for beam loss and background control in Future Linear Colliders (FLC) especially near the interaction point (IP). The Accelerator Test Facility 2 (ATF2) \cite{atf22005atf2,atf22005atf2_2} at KEK in Japan is a scaled down Final Focus System (FFS) prototype of the FLC with the aim to demonstrate nanometer level focusing based on a scheme of local chromaticity correction \cite{raimondi2001novel}. The ATF2 layout is shown in Fig.~\ref{fig:ATF2}. At ATF2, beam halo hitting on the beam pipe in the final doublet and after the IP can generate a large amount of background (through bremsstrahlung) for the measurement of the nanometre beam size using the laser interferometer beam size monitor (IPBSM, also known as Shintake monitor) \cite{shintake1992proposal}. Dedicated collimators are proposed to collimate the beam halo upstream \cite{Nuria_IPAC2015}. Since the design of the collimators strongly relies on the transverse beam halo distribution, which is currently unknown for the ATF2 beam line, dedicated measurements are essential. Such measurements will also allow us to distinguish different sources of beam halo, several of which have been studied analytically~\cite{Dou2014analytical}.

One important issue for beam halo measurements is to reach a large dynamic range, since beam core measurements are required for the proper normalization and parameterization of the beam halo distribution. Beam halo measurements using wire scanners in the old extraction (EXT) line of ATF reached a dynamic range of $\sim10^4$~\cite{suehara2008design}. Recent wire scanner measurements in the present ATF2 beam line however only achieved a dynamic range of $\sim10^3$ due to less favourable background conditions~\cite{liu2014beam}. Single crystalline Chemical Vapor-Deposition (sCVD) diamond sensors (DS) are not only sensitive to a single electron but have also been tested to have a linear response up to $10^7$ electrons with a 50 $\Omega$ readout system~\cite{liu2014status}. Therefore, two sCVD \textit{in vacuum} Diamond Sensors (DSv) with four strips each have been developed with the aim of achieving large dynamic range ($ \sim10^6$) for simultaneous measurements of beam core and beam halo. These DSv were installed after the BDUMP bending magnet located downstream of the IP (see Fig.~\ref{fig:ATF2}). This location not only allows us to investigate the beam halo transverse distributions but also provides us the possibility to probe the Compton recoil electrons generated at the IP~\cite{Thesis_Shan}.

For the study of diamond sensor response under very high beam intensity ($> 10^7$ electrons), knowledge of charge carrier transport parameters (mobility, saturation velocity, lifetime etc.) is required~\cite{slava}. Therefore, before the implementation of the DSv at ATF2, characterization of a sCVD DS sample fabricated by CIVIDEC \cite{CIVIDEC}, with a similar diamond sample as used for the DSv, was carried out using the Transient Current Technique (TCT)~\cite{pernegger2005charge} and using the electron beam at PHIL (low energy ($<10~$MeV) photo-injector at LAL)~\cite{alves2013phil}. 

In this report, we present the diamond sensor characterization using TCT and electron beam, followed by the design of DSv for ATF2 and its performance studies in the first operation at ATF2. First beam core and beam halo measurements results using DSv will also be presented.

\section{Characterization of diamond sensor using TCT}
Characterization of diamond sensor was performed with an $\alpha$ source. The tested diamond sample has a dimension of $4.5~mm \times 4.5~mm \times 500~\mu m$. As estimated penetration depth of $\alpha$ particles is less than 20 $\mu$m in diamond, which is small compared to the 500 $\mu$m thickness DS, the electron hole pairs generated in this process can be considered as a thin layer of charge. By changing the polarity of the bias voltage, the transport parameters of electrons and holes can be studied separately. Therefore, measurements using $\alpha$ sources are considered for detector characterization using TCT. 

\begin{figure*}[!tbh]
    \centering
    \includegraphics*[width=40mm]{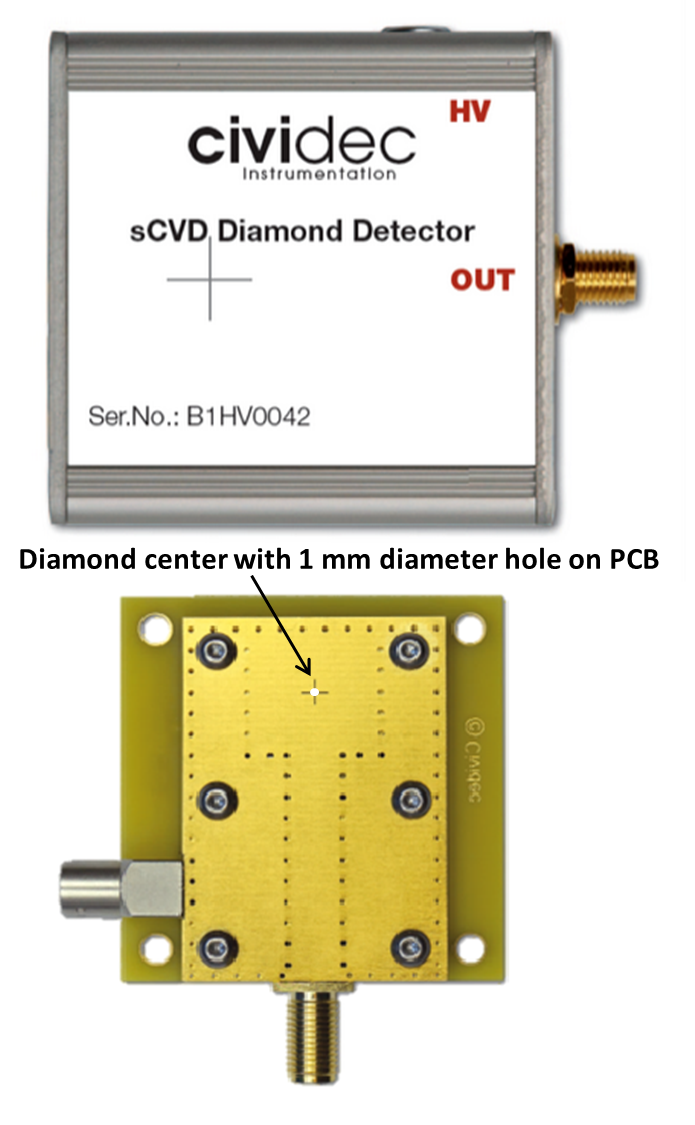}
    \includegraphics*[width=80mm]{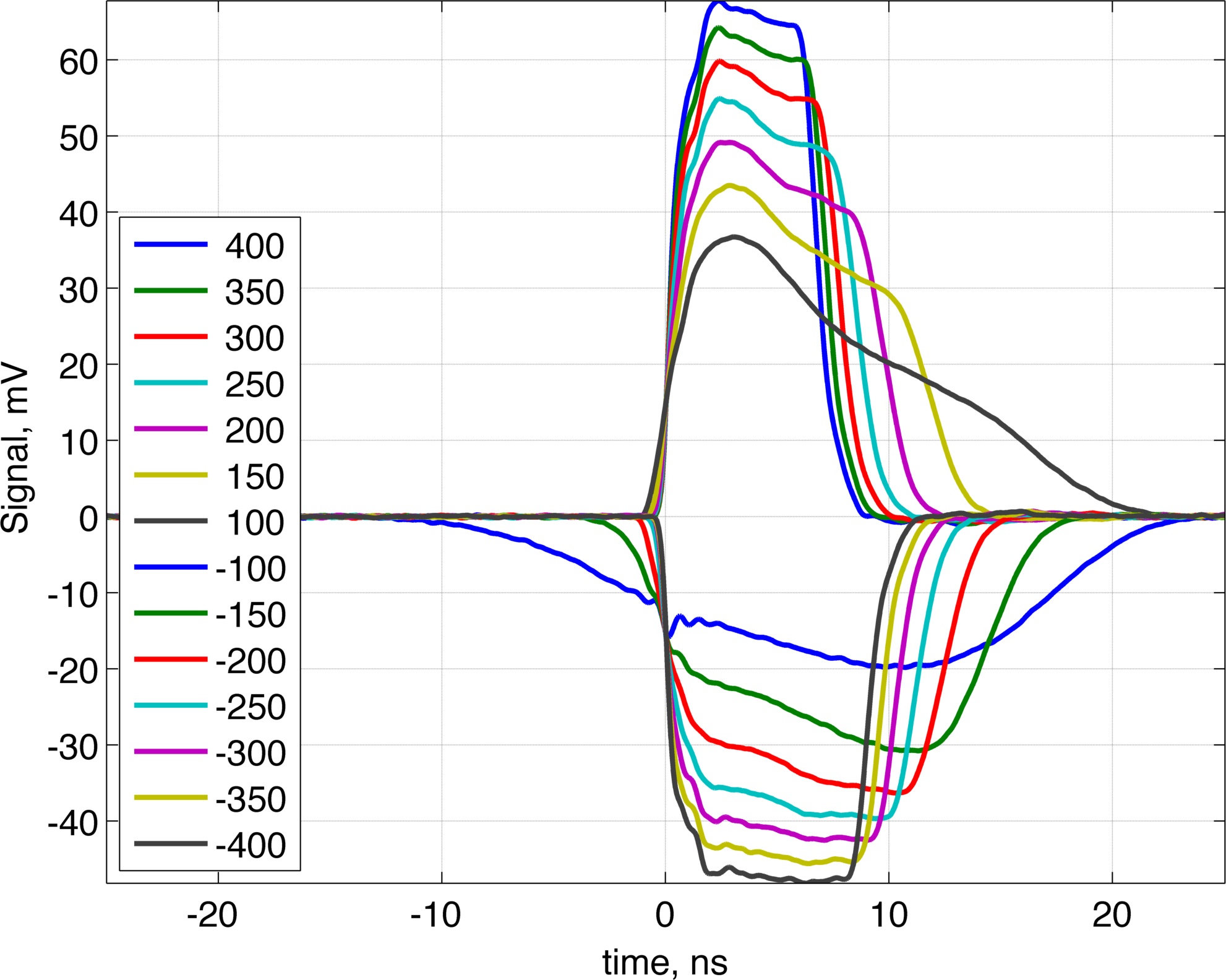}    
     \caption{Left: DS sample fabricated by CIVIDEC company with aluminum box (upper) and PCB cover (lower) for RF sheilding (the cross indicates the center of the diamond); Right: signal from collected electrons (negative pulse) and holes (positive pulse).}
    \label{fig:alpha}
\end{figure*}

For the $\alpha$ particles injection, an $^{241}Am$ source with an activity of 3.9 kBq was mounted with a distance of 2 mm on the top center of the diamond, where a 1 mm diameter hole was specially made on the PCB for this test (see Fig.~\ref{fig:alpha} lower left). The $^{241}Am$ source is  3~mm thick and has a diameter of 25~mm. The energy of the $\alpha$ particles emitted from $^{241}Am$ source is around 5.4 MeV.

Fig.~\ref{fig:alpha} (right) shows pulse shapes from electrons and holes by averaging 1000 events for data taken at different bias voltages. The bias voltage was changed from -400 V to 400 V in steps of 50 V. The settling time between each step is 600 s. A 40 dB C2 broadband (1 MHz - 2 GHz) current amplifier \cite{CIVIDEC} was used for the measurements.  Since the noise level of the C2 amplifier is 2.5 mV ($\sim10^4 e^-$), a self-trigger level of $\pm$10 mV was applied for positive and negative bias voltage separately. Due to this reason, our measurements is limited to absolute values of bias voltage above 100 V.

From Fig.~\ref{fig:alpha} it can be seen that pulse widths increase as the bias voltage decreases, which indicates a slow down of drift velocity ($v_{dr}$) due to the decrease of the electric field. However, the rise time of the current signal, which is dominated by the time constant of the electronic circuit remains almost unchanged\footnote{With around 2 pF of capacitance for the DS, the time constant is around 0.1 ns.}. After this fast rising, the signal level should keep constant during the drift of charge-carriers. However, as there is a net effective space charge in the diamond bulk, we observe an increasing current for electrons and a decreasing current for holes, which indicates a negative space charge in the bulk. This space charge effect is explained in more detail in Ref.~\cite{pernegger2005charge}.

\subsection{Charge collection efficiency}
\begin{figure*}[htb]
    \centering
     \begin{subfigure}[b]{0.5\textwidth}
     \includegraphics*[width=75mm]{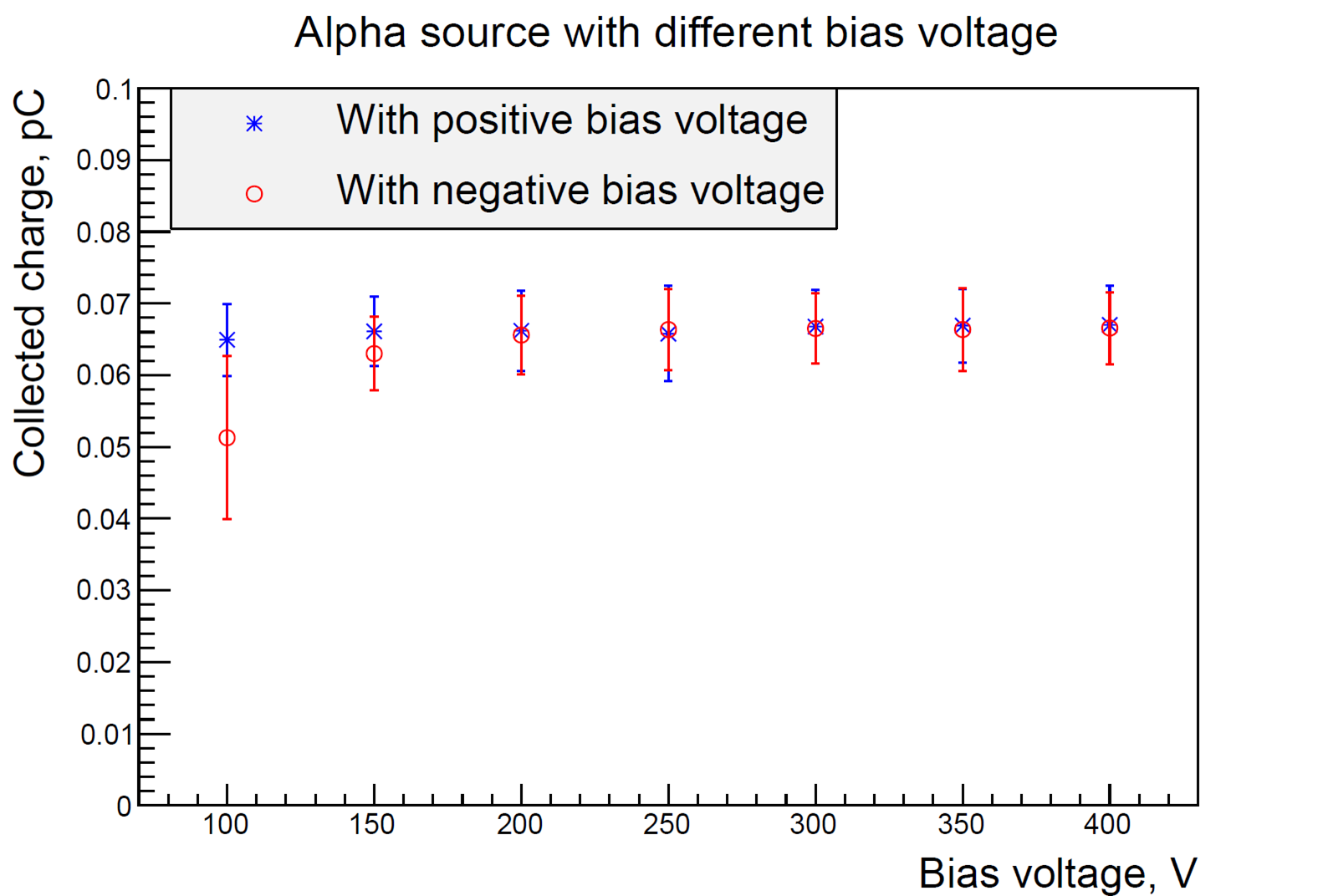}
     \caption{}  
    \end{subfigure}%
      \begin{subfigure}[b]{0.5\textwidth}
     \includegraphics*[width=75mm]{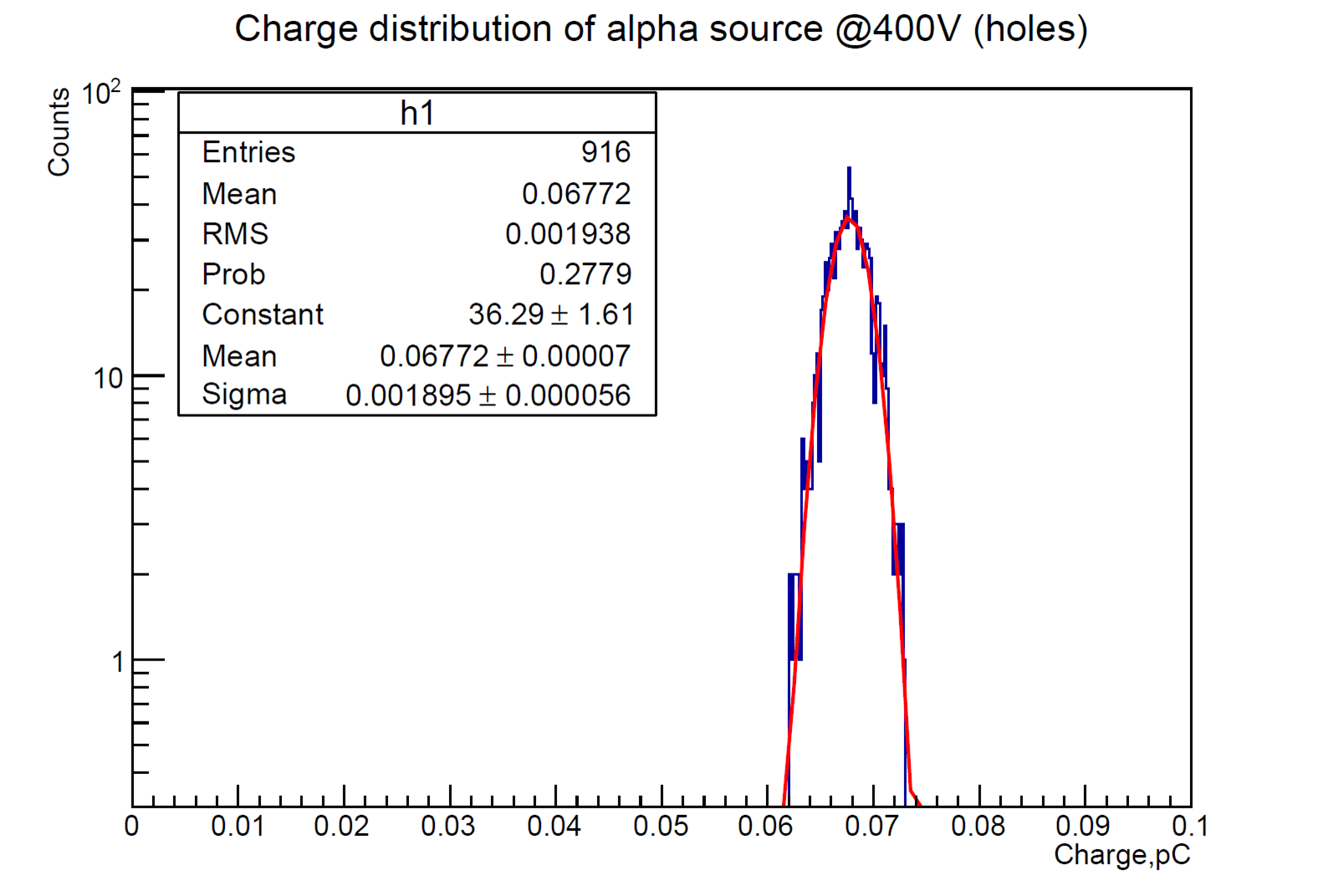}
         \caption{}  
    \end{subfigure}
    \caption{ (a) collected charge as a function of bias voltage for electrons (red) and holes (blue); (b) histogram of charge collected from the alpha source at 400V.}
    \label{fig:CCE_alpha}
\end{figure*}

The charge collection efficiency (CCE) is defined as the ratio between the total generated charge ($Q_{gen}$) and the charge collected ($Q_{coll}$) at the electrodes: $CCE=Q_{coll}/Q_{gen}$. Fig.~\ref{fig:CCE_alpha} shows $Q_{coll}$ as a function of applied bias voltage (a) and the distribution of $Q_{coll}$ for the selected events at 400 V (b) . Although at each voltage 1000 events have been taken, there is a small fraction of ``bad" events which may be caused by the RF pick-up in the outside environment. Therefore, signal selection criteria were set requiring all events to have a collected charge ($Q_{coll}$) of more than 20 fC and charge collection time (FWHM, $t_{FWHM}$) of more than 6 ns. The collected charge gets saturated after applying more than $\sim$150 V for the collection of both holes and electrons. The maximum collected charge at 400 V is 67.7$\pm$1.9 fC for holes and 66.6$\pm$5.0 fC for electrons. As the mean ionization energy is around 13 eV in diamond, the expected charge from the $\alpha$ source is around 66.5 fC. Therefore, the CCE of this diamond sample is $100\%$ for both electrons and holes for voltages larger than 150 V.

\subsection{Charge-carrier transport parameters}
During the charge collection process, the charge-carriers have a certain probability to get trapped or recombined due to different kinds of defects inside the diamond bulk. The lifetime of the charge-carriers ($\tau _{e,h}$) is defined as the time for the total generated charge $Q_0$ to decrease by a factor of $e$. It can be derived from the measurements using $\alpha$ sources. The charge-carrier mobility and saturation velocity can also be determined using the TCT method.
  
The drift velocity of the charge-carriers in the diamond as a function of electric field $E$ can be described using the empirical formula \cite{pernegger2005charge,li1993modeling}:
\begin{equation}
v_{dr}=\frac{\mu _{0_{e,h}}\cdot E}{1+\frac{\mu _{0_{e,h}}}{v_{sat_{e,h}}}\cdot E}
\label{equa:drift}
\end{equation} 
where $\mu _{0_{e,h}}$ is the low-field mobility for electrons and holes and $v_{sat_{e,h}}$ 
the saturation velocities for electrons and holes.

In the measurements, the drift velocity can be defined as $v_{dr}(E)=d/t_{tr}$, where d is the diamond thickness and $t_{tr}$ is the transit time, taken as the FWHM of the signals from Fig.~\ref{fig:alpha} as in Ref.~\cite{pomorski2006development}\footnote{The transit time $t_{tr}$ was taken as the total transit time $t_{tr}$=$t_e$-$t_s$ in Ref.~\cite{pernegger2005charge}, where $t_s$ and $t_e$ are the signal start and end times, respectively.}. 

\begin{figure*}[pht]
    \centering
   \begin{subfigure}[b]{0.5\textwidth}
    \includegraphics*[width=75mm]{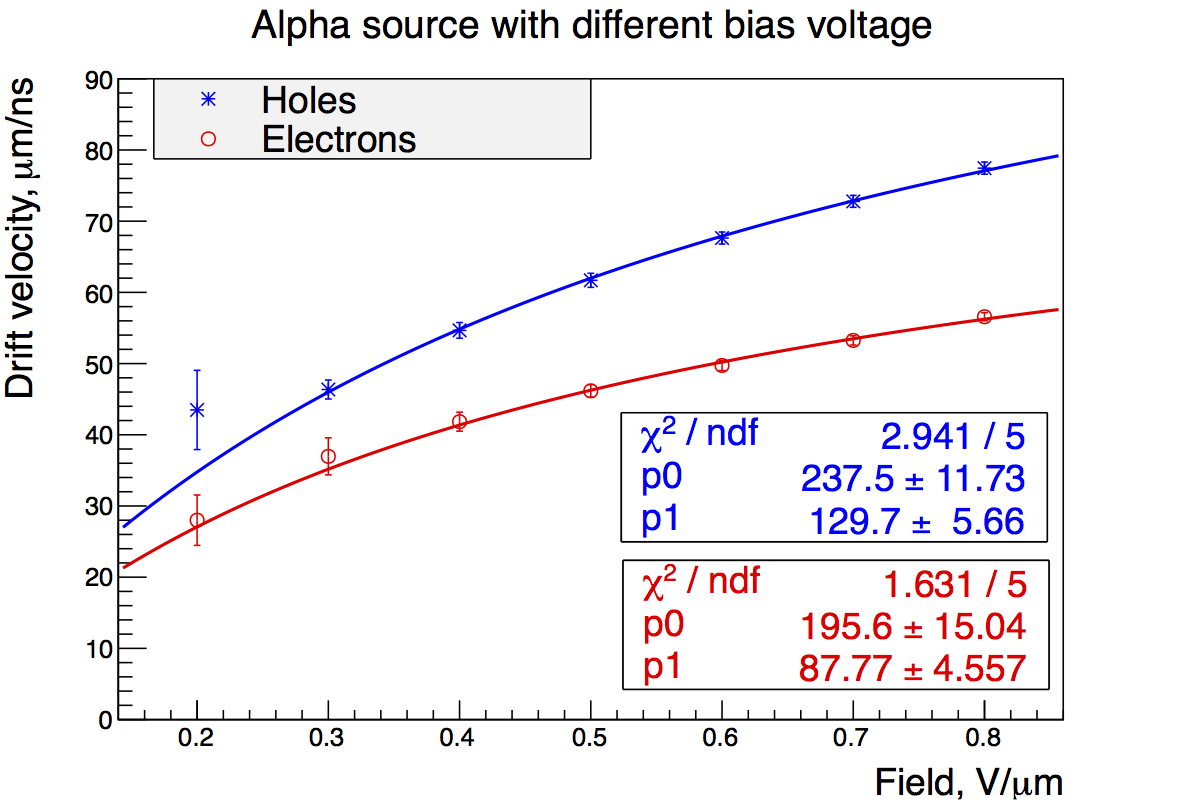}
         \caption{}  
    \end{subfigure}%
   \begin{subfigure}[b]{0.5\textwidth}    
        \includegraphics*[width=75mm]{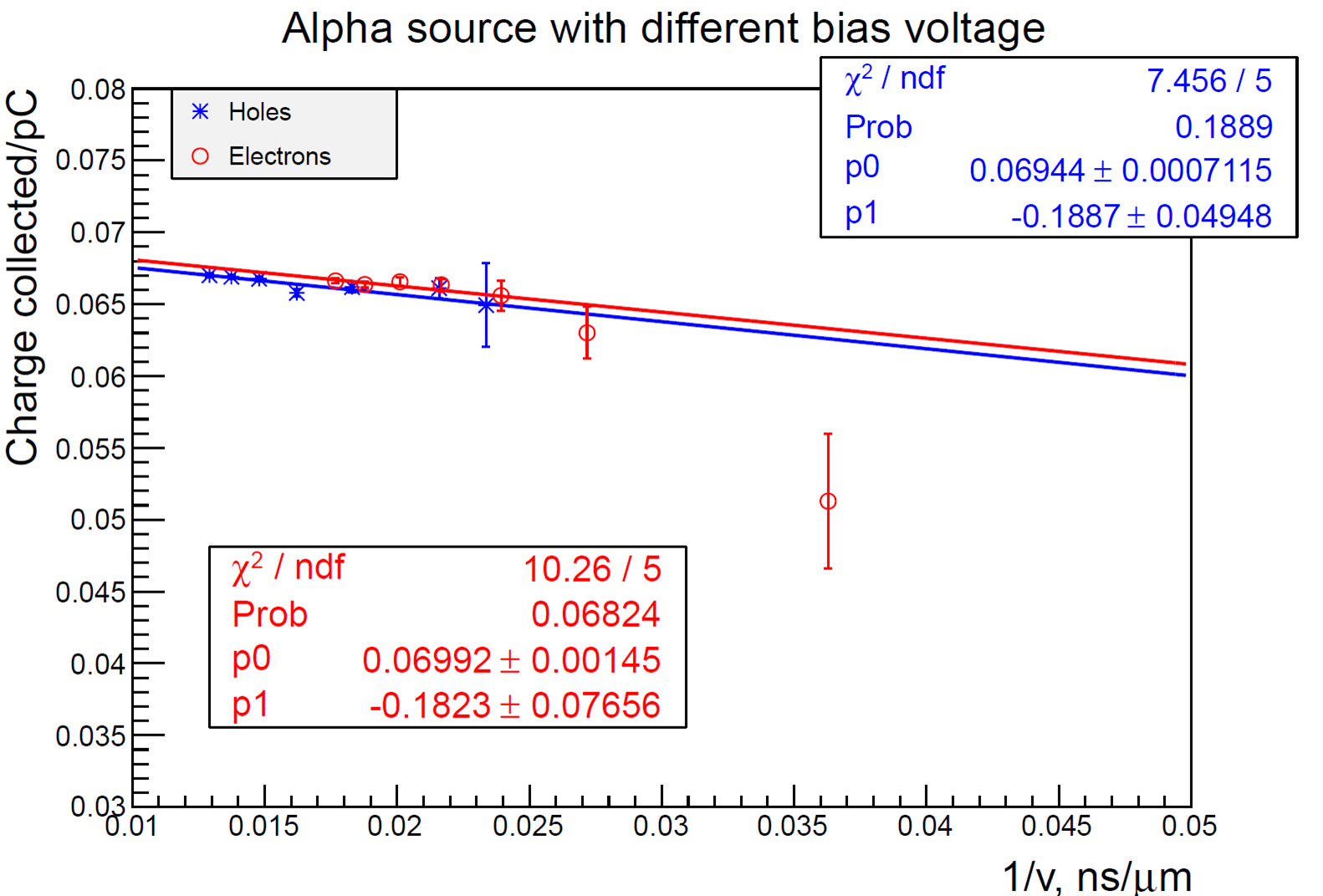}
                 \caption{}  
    \end{subfigure}
    \caption{(a) drift velocity as a function of electric field: $v_{dr}=\frac{p_{0}\cdot E}{1+\frac{p_{0}}{p_{1}}\cdot E}$; (b) collected charge as a function of inverse drift velocity: $Q_{e,h}=p_{0}+p_{1}\frac{1}{v_{dr}}$.}
    \label{fig:V_drift}
\end{figure*}

Fig.~\ref{fig:V_drift} (a) shows the drift velocity as a function of applied electric field. The curves for electrons and holes were fitted according to Eq.~\ref{equa:drift}. From the fit parameters of the curves, the mobility and saturation velocity of electrons and holes can be obtained as described below.

The total induced charge can be defined as the integration of the measured current pulse:
\begin{equation}
\label{equa:integration}
Q_{e,h}(V)= \int_{0}^{t_{tr}} \! i_{m_{e,h}}(V,t) \, \mathrm{d}t 
\end{equation} 
and assuming that the lifetimes of electrons and holes, $\tau _{e,h}$, are much larger than $t_{tr}$, we can get a solution as: 
\begin{equation}
Q_{e,h}(V)= Q_{0_{e,h}} \left( \frac{\tau _{e,h}}{t_{tr}} \right) \left( 1-e^{-\frac{t_{tr}}{\tau_{e,h}}} \right)\approx Q_{0_{e,h}}(1-\frac{t_{tr}}{2\tau _{e,h}})
\label{equa:lifetime}
\end{equation} 
Since $t_{tr}$ can be expressed in terms of the inverse drift velocity as $d/v_{dr}$, the measured charge $Q_{e,h}(V)$ can be plotted as a function of the inverse drift velocity $1/v_{dr}$ as shown in Fig.~\ref{fig:V_drift} (b). The data were fitted linearly to extrapolate the lifetime using Eq.~\ref{equa:lifetime}. 

The extrapolated values of the low-field mobility ($\mu _0$), saturation velocity ($v_{sat}$), lifetime ($\tau$) and the calculated maximum charge collection distance ($d_{c_{max}} = v_{sat} \times \tau $) of electrons and holes are shown in Table ~\ref{table:lifetime2}. It can be seen that $d_{c_{max}} \gg 500 \mu$m, the thickness of the DS. 

\begin{table}[htp]
\centering
\begin{tabular}{cccccc} 
{}&{}&{\bf $\mu _0$, [$cm^2$/Vs]} & {\bf $v _{sat}$, [$\mu$m/ns]} & {\bf $\tau $, [ns]} & {\bf $d_{c_{max}} $, [mm]}\\ 
\hline 
 \bf{Electron} &measured & 1956$\pm$150.4 & 87.77$\pm$4.56  & 92.00$\pm$25.07 & 8.07$\pm$2.24\\
\bf{} &reported & 1300-4500~\cite{Thesis_Christina_Weiss} & 96-270~\cite{pernegger2005charge,neves2001properties} &40-321 \cite{pernegger2005charge,pomorski2006development}  &-\\
\bf {Holes}& measured & 2375$\pm$117.3 & 129.7$\pm$5.66  & 95.89$\pm$42.24 & 12.44$\pm$5.50\\ 
\bf {} &reported  & 2050-3800~\cite{Thesis_Christina_Weiss}& 85-140~\cite{pomorski2006development,wort2008diamond}  & 40-968 & -\\
\hline
\end{tabular}
\caption{Charge transport parameters for the measured DS sample and ranges of values reported.}
\label{table:lifetime2}
\end{table}

\section{Characterization of diamond sensor using electron beam}
After the characterization of the diamond sensor using TCT, characterization using electron beam were performed at PHIL to study the waveform shapes for different beam intensities and to measure the linearity of the diamond sensor response. 

\begin{figure*}[!tbh]
    \centering
    \includegraphics*[width=100mm]{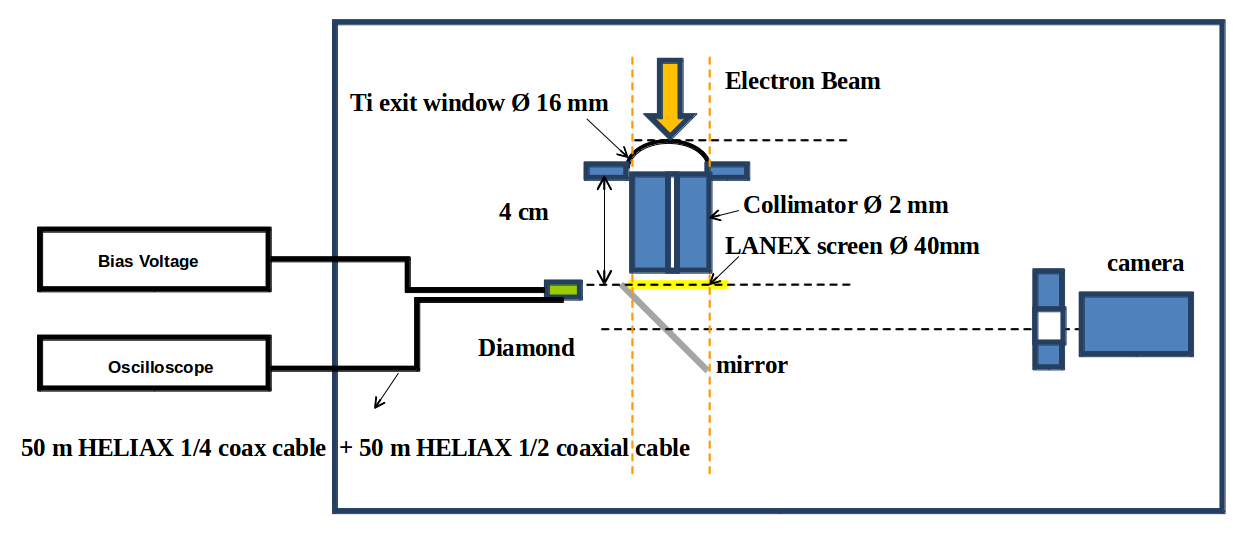}
     \caption{Experimental setup at PHIL.}
    \label{fig:PHIL}
\end{figure*}

PHIL is a 5 m long photoinjector beamline at LAL, which provides an electron beam with low energy ($<5$ MeV), low emittance ($10 \pi \cdot mm\cdot mrad$) and short pulse (7 ps). The bunch frequency at PHIL is 5 Hz. The electron beam charge can be varied from less than 1 pC to 2.2 nC at the exit window by changing the laser density shining the Mg cathode. Fig.~\ref{fig:PHIL} shows the experimental setup for the tests at PHIL. A 4 cm long Al collimator with 2 mm diameter was used to collimate the beam after the exit window. A LANEX (R) screen was used to measure the charge, it is capable to measure beam charges as small as 33$\pm$11 fC~\cite{vinatiermeasurement}. The diamond detector was mounted next to the LANEX screen on a transverse movable stage after the collimator. A 50 $\Omega$ feed through terminator was used to enable the 1 M$\Omega$ coupling on the scope, and the signals were viewed with the 50 $\Omega$ termination by the scope.

\subsection*{Pulse waveforms}
\begin{figure*}[!tbh]
    \centering
       \begin{subfigure}[b]{0.5\textwidth}
    \includegraphics*[width=75mm]{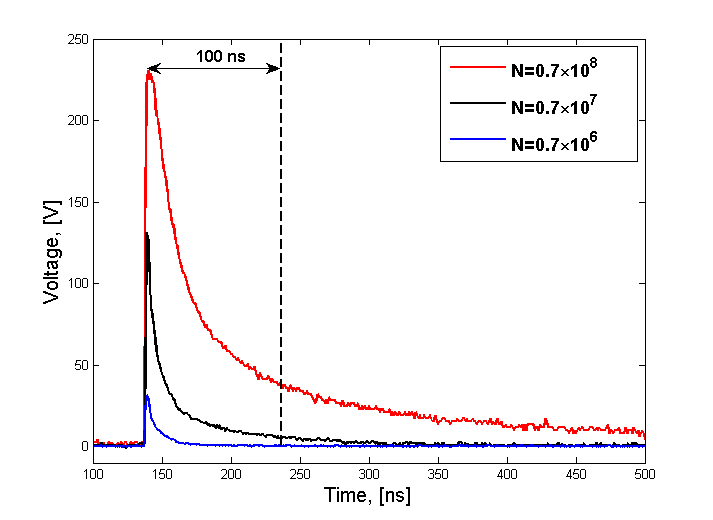}
             \caption{}  
    \end{subfigure}%
           \begin{subfigure}[b]{0.5\textwidth}
    \includegraphics*[width=75mm]{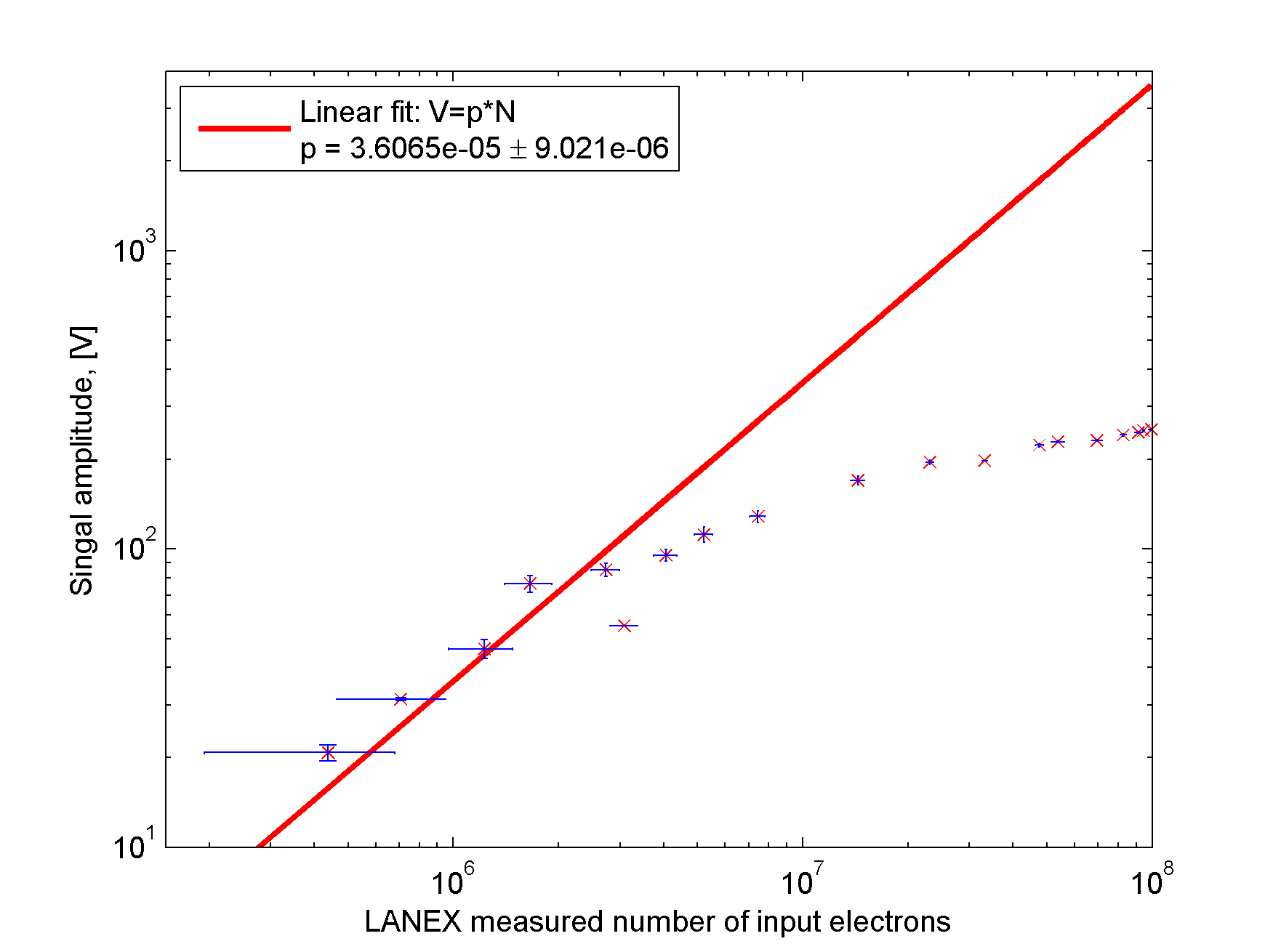}
             \caption{}  
    \end{subfigure}%
     \caption{Signal pulse form (a) and signal amplitude (b) for different injected beam intensities.}
    \label{fig:pulse}
\end{figure*}

Fig.~\ref{fig:pulse} (a) shows examples of pulse shapes for different beam intensities. We can see that the pulse width enlarges with the beam intensity. This can be explained by the voltage drop on the 50 $\Omega$ termination. Immediately after the passage of a high intensity beam, a sudden rise of current ($I_{induced}$) is induced due to the fast movement of electrons and holes in the externally applied electric field. Consequently, a large voltage drop ($V_{drop}$) is generated on the 50 $\Omega$ termination as $V_{drop}=I_{induced}\times 50~\Omega$. The voltage drop reduces the effective bias voltage on the DS ($V_{eff}$):
\begin{equation}
V_{eff}= V_{bias}-V_{drop} 
\end{equation}
where $V_{bias}$ is the bias voltage (400 V in our case). When a large voltage drop occurs, the drift velocities of electrons and holes are decreased, the charge collection time is extended. $V_{drop}$ can be read out directly from the waveforms on the scope and it changes together with the current. The maximum voltage drop correspond to the signal amplitude.

\subsection*{Dependence of signal amplitude on intensity}
The dependence of signal amplitude on intensity is shown in Fig.~\ref{fig:pulse} (b)\footnote{During the measurement, 10 images were taken by the LANEX screen at each intensity to get the averaged charge and beam size for calculating the number of intercepted electrons. It can be seen that the fluctuation of the measurements is much more significant for lower intensity than for higher intensity.}. The signal amplitude can be calculated as $V_{max}$=$I_{max}\times50~\Omega$, where $I_{max}$ is the maximum induced current, when the charge carriers move with their saturation velcocities. According to the Shockley-Ramo formula~\cite{shockley1938currents,ramo1939currents}, the maximum instantaneous current generated by $N$ input electrons can be calculated as:
\begin{equation}
\label{eq:Shockley-Ramo}
I_{max}=E_v\cdot Q_{total}\cdot v_{sat}~~ \mathrm{with}~~Q_{total}=N\cdot Q_{MIP} 
\end{equation}
where $E_v$ is the normalized electric field, $Q_{total}$ is the generated total charge, $v_{sat}$ is the saturation velocity and $Q_{MIP}$ is the charge generated by 1 Minimum Ionisation Particle (MIP)\footnote{The average minimum ionizing signal per 100 $\mu m$ is 3600 electrons. For 500 $\mu m$ thick diamond, the charge generated by 1 MIP is $\sim$2.88 fC.}. 

From Fig.~\ref{fig:pulse} (b) it can be seen that the pulse amplitude rises almost linear for less than 2$\times 10^6$ electrons and then it starts to get saturated. By fitting the linear region in the plot, a saturation velocity velocity of $125.23\pm 31.25~\mu m/ns$ can be extrapolated, which is consistent with the measured value using TCT (see Table~\ref{table:lifetime2}). This means that the drift velocity of the charge carriers are not much affected by the voltage drop up to $\sim10^6$ electrons. However, after an input of more than $10^6$ electrons, due to large voltage drop, the charge carriers can no longer reach their saturation velocities\footnote{In this case, the signal amplitude starts to get saturated and the calculated $V_{max}$ using the $I_{max}$ from Eq.~\ref{eq:Shockley-Ramo} gives unphysical results (more than the bias voltage of 400 V) for $N>10^7$.}, and much longer time is required for charge collection. Nevertheless, as long as the charge collection time is smaller than the lifetime of the electrons and holes, a linear charge collection response can be expected.

\subsection*{Dependence of collected charge on intensity}
\begin{figure*}[!hbt]
    \centering
    \includegraphics*[width=80mm]{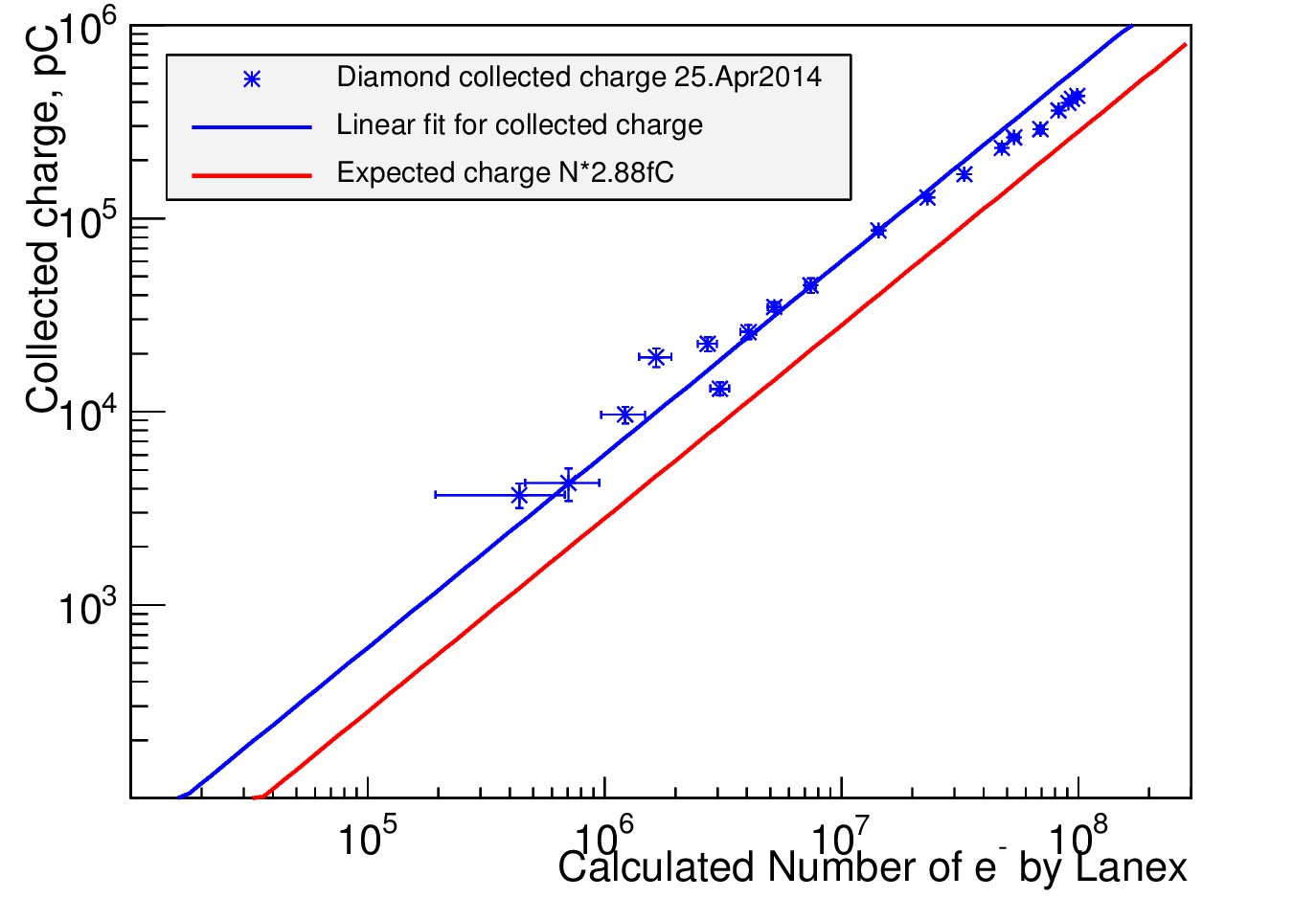}
     \caption{Collected charge as a function of number of input electrons. The blue line is the linear fit of the data and the red line is the expected signal.}
    \label{fig:charge}
\end{figure*}

Fig.~\ref{fig:charge} shows the dependence of collected charge on intensity. In Fig.~\ref{fig:charge}, the expected signal (red) was calculated as the charge generated by 1 MIP (2.88 fC) multiplied by the number of electrons derived from the charge measured by the LANEX. A factor two difference was observed between the linear fit (blue line) and the expected signal (red line). This may be due to some bias in the calibration procedure of the LANEX screen, in which case the diamond sensor could be used as absolute reference for the LANEX, or it could be due to some other systematic effect from relative acceptances or alignment of the two devices. In either case, this discrepancy is not a major concern for a first estimation of the non-linear response level of the DS. A linear response up to $\sim10^7$ electrons can be observed from Fig.~\ref{fig:charge}. A hint of saturation in the collected charge (decrease in CCE) can also be seen for input charges larger than $10^7$ electrons. 

The saturation of charge collection can be explained by comparing the charge collection time with the lifetime of the charge carriers. In Fig.~\ref{fig:pulse} (a), it can be seen that the charge collection time for $N \approx 10^7$ electrons is comparable with the lifetime ($\tau \approx 100~ns$) obtained using the TCT (see Table \ref{table:lifetime2}), however, for $N \approx 10^8$, it is much longer. Therefore, not all the charge can be collected $N \approx 10^8$ (CCE$<100\%$). Such behaviour was already shown in Fig.~\ref{fig:CCE_alpha} (b) for bias voltages explicitely lowered below 150 V. From Fig.~\ref{fig:pulse} (b) it can be seen that the voltage drop for $N=10^8$ can reach $\sim300$~V. However, the decrease of CCE caused by such a voltage drop may be mitigated by adding a smaller resistor in parallel to the 50~$\Omega$, as a current divider\footnote{Alternative way of avoiding the signal saturation is either to use thinner diamond (i.e. decrease $Q_{MIP}$) or to increase the electric field (i.e. increase $V_{bias}$). However, the maximum bias voltage can be applied to the DS is limited by the leakage current level, which depends on the quality of the diamond.}. This has been successfully demonstrated at the Beam Test Facility (BTF) in Frascati, Italy, using a 100 $\mu m$ thick polycrystalline diamond (pCVD). In their experiments, a linear response up to $10^9$ electrons has been measured by adding a 1~$\Omega$ shunt resistor~\cite{steinresponse}. 

Besides the voltage drop, the space-charge within the diamond, resulting from the separation of electrons and holes, is another reason for the increase of charge collection time. Modelling of charge collection processes taking into account both the voltage drop and space charge effects is explained in more detail in Ref.~\cite{slava}.

\section{Design of the \textit{in vacuum} diamond sensor scanner for ATF2}
\begin{figure*}[!tbh]
    \centering
    \includegraphics*[width=120mm]{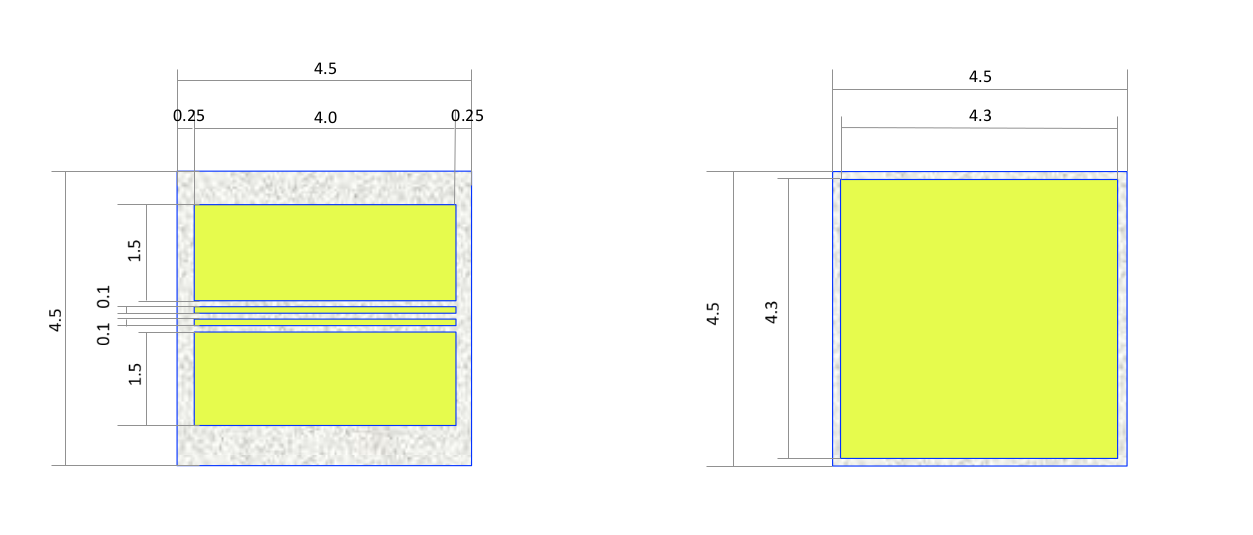}
     \caption{DSv with 4 strips:topside (left) and bottom side (right).}
    \label{fig:DS_4strips}
\end{figure*}

For the measurements at ATF2, a large signal range from $10^2~e^{-}/mm^2$ for Compton and $10^4~e^{-}/mm^2$ for beam halo to $10^8~e^{-}/mm^2$ for beam core is required~\cite{liu2013development}. As in first approximation the signal strength is proportional to the metallised surface of the electrode on the diamond, a geometry with four strips was adopted to help cover this large dynamic range (see Fig. \ref{fig:DS_4strips}). The two strips designed for beam halo scanning are on the two outer sides, with dimensions of 1.5 mm$\times$ 4 mm, and the other two designed for beam core scan are in the center, with dimensions of  0.1 mm$\times$4 mm. 

\subsection{PCB and electronic circuit design}
\begin{figure*}[htb]
    \centering
    \includegraphics*[width=100mm]{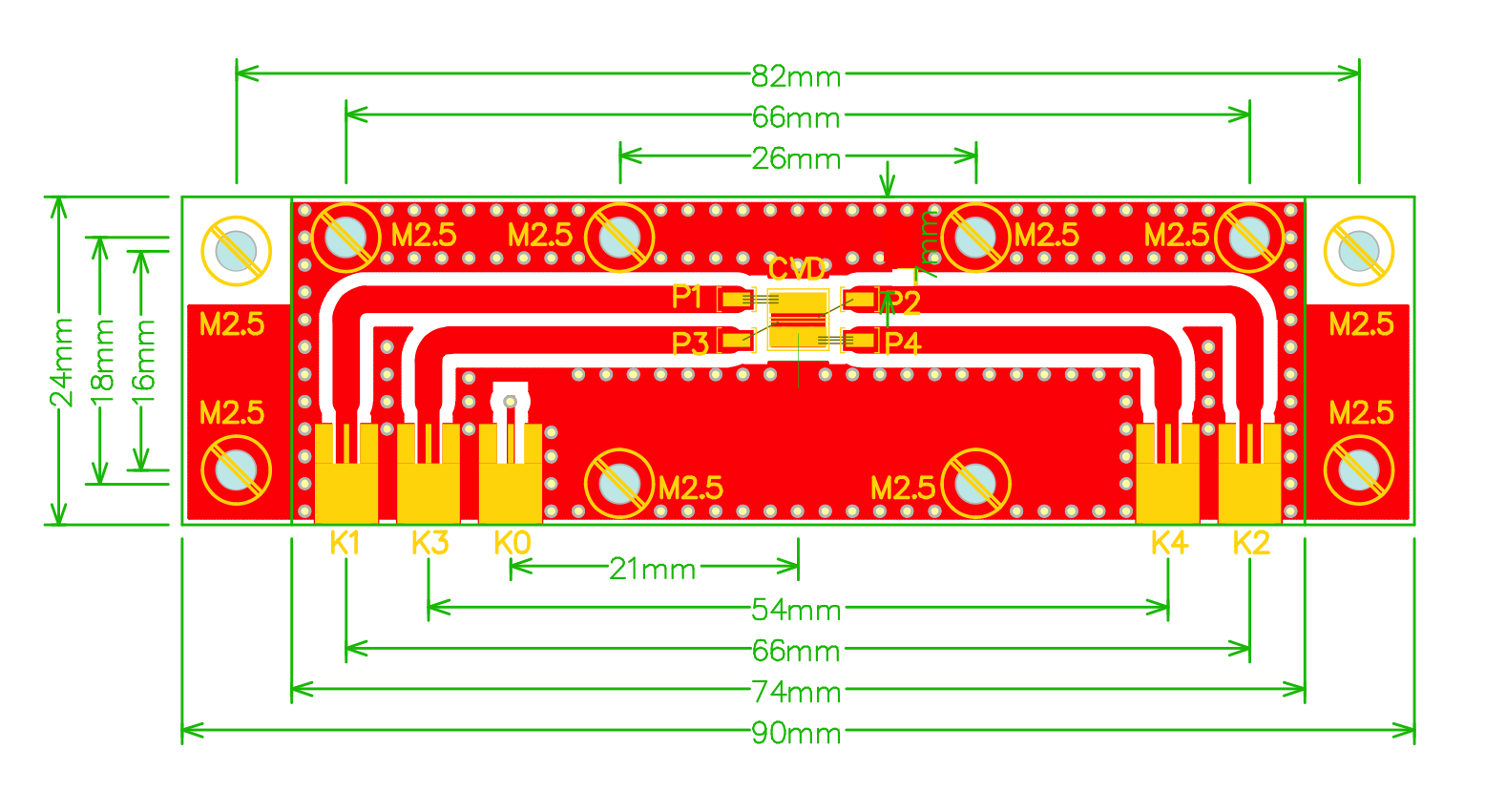}
    \includegraphics*[width=100mm]{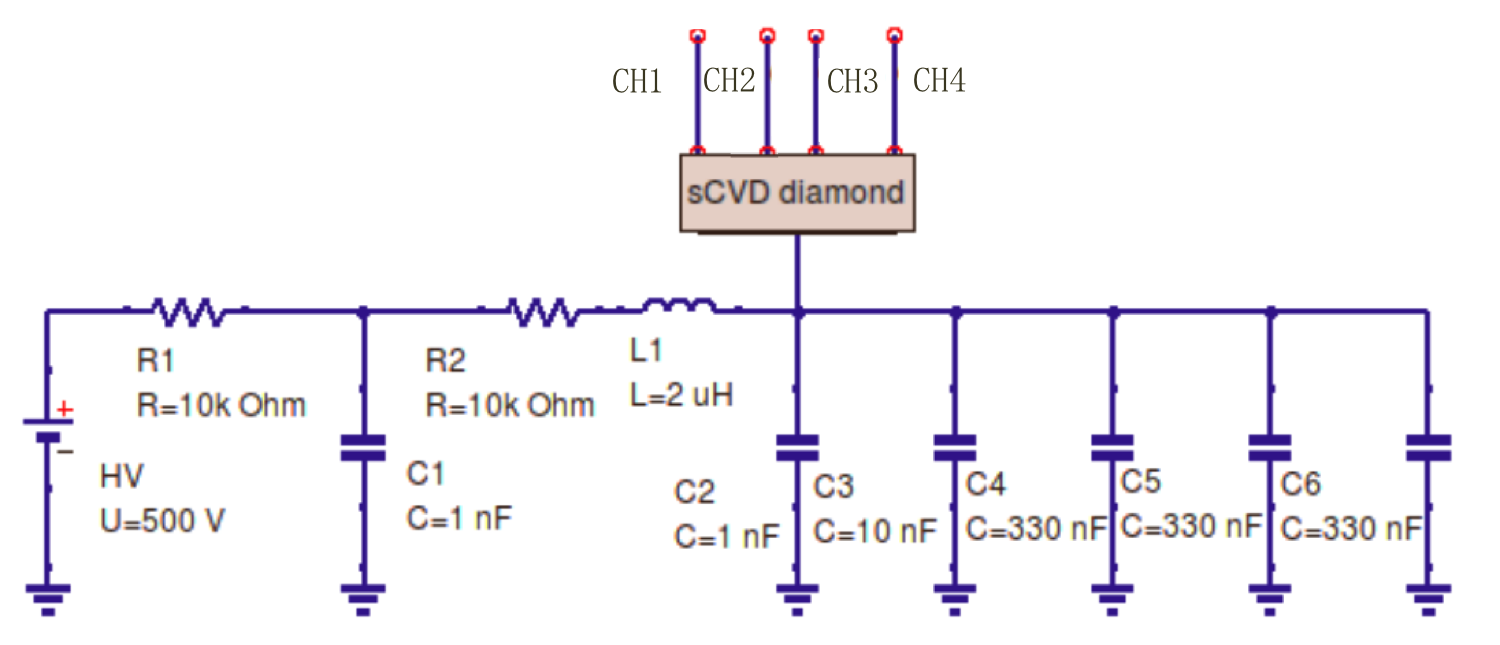}
     \caption{Layout of the PCB (top) and electronic circuit (bottom) for DSv.}
    \label{fig:PCB}
\end{figure*}

For the \textit{in vacuum} application a ceramic PCB is used. The ceramic PCB uses silver-platinum conductor produced in thick-film technology. The PCB and the electrical circuit for the diamond detector are shown in Fig. \ref{fig:PCB}. A low pass filter together with charging capacitors are mounted on the backside of the ceramic PCB. The parameters of this circuit were set based on the following considerations:

\begin{itemize}

  \item The cut-off frequency for the high voltage power supply should be as low as possible to maintain the stability;
  \item The amount of charge stored on the capacitors should be large enough for measurements of large beam intensities, up to $\sim1~\mu$C on the narrow strips;
  \item The charging time constant should be small to enable separating successive bunches, the bunch repetition rate being 5 Hz at PHIL and 3 Hz at ATF2; 

\end{itemize}

The simulated cut-off frequency for the present circuit is 8 Hz with 500 $\mu$C maximum stored charge at 500 V and the charging time constant is 44 ms. These values can fulfill our requirements listed above.

\subsection{Mechanical design}
\begin{figure}[!htb]
   \centering
   \includegraphics*[width=120mm]{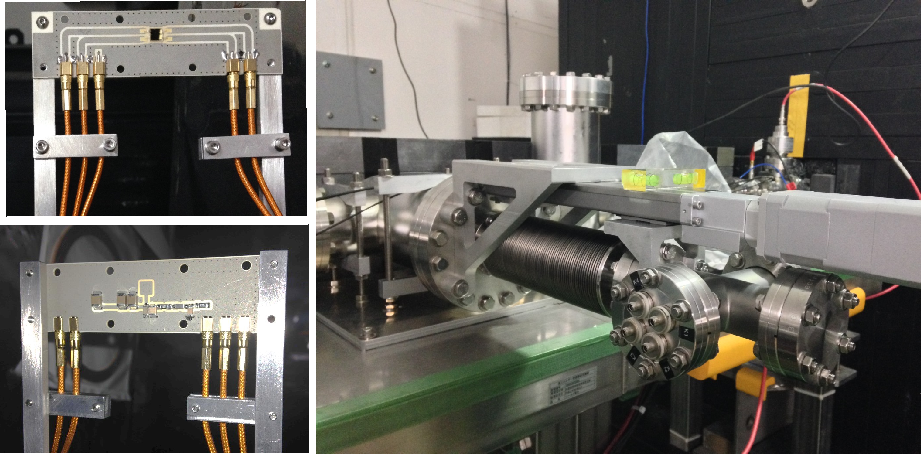}
   \caption{Layout of the DSv with the front side (upper left) and back side (lower left) of the PCB and the motor used for the horizontal scan (right).}
   \label{fig:installation}
\end{figure}

The diamond sensor is designed to be installed in vacuum with a holder to scan the beam and beam halo. The whole setup is 0.8 m long and it can be oriented either horizontally or vertically to scan in different axes. Fig.~\ref{fig:installation} shows the layout of the DSv with the PCB on the holder and the motor used for the horizontal scan. The vacuum chamber with an inner diameter of 72.1 mm is connected to the beam pipe (diameter of 63 mm) and installed after the BDUMP bending magnet at ATF2 via two DN63CF flanges. Since the mechanical design is compatible for PHIL and ATF2, it was first installed and tested at PHIL in October 2014, as a preparation for the implementation at ATF2.

\section{First operation of \textit{in vacuum} diamond sensor scanner at ATF2}

\subsection{Experimental setup}

The horizontal and vertical DSv scanners were installed at ATF2 in November 2014 and March 2015, respectively (see Fig.~\ref{fig:ATF2} for the installation location). The horizontal DSv is inserted from the low energy (LE) side to the high energy (HE) side and the vertical DSv is inserted from the upper side to the lower side of the beam.

For the reason of radiation safety, the oscilloscope and other readout electronics were located in the IP laser room outside the tunnel. Therefore, in order to collect the signal from the DSv in the post-IP region, LDF1-50 1/4 inch thick Heliax coaxial cables (over a distance of 20 m) were used. A Keithley 2410 sourcemeter, which can supply bias voltages of $\pm$1000 V, was used for the bias voltage supply. The whole data acquisition system is connected via Ethernet cables to the PC located in the control room. On the PC, a Matlab program \cite{guide1998mathworks} is installed and used for the data taking. The oscilloscope we used is an Agilent oscilloscope with a sampling rate of 4 GS/s and an analogue bandwidth of 1 GHz. The stepper motor we used for DSv scanning is driven by a ball bearing spindle of 2 mm pitch for one revolution. The stepping motor performs one rotation by 200 steps (stepper angle 1.8$\pm$5\%). Combining the ball bearing spindle and the precise stepping motor, the resolution of the steps is 10 $\mu$m\footnote{Since the scan speed depends on the moving step precision, it was set to 50 $\mu$m in the our program to enable relatively fast scanning.} and the reproducibility can be as good as 3$\mu$m. The maximum scan distance is 110 mm, which can cover the whole region of beam halo scan\footnote{The effective scan range is 63 mm (vacuum beampipe diameter), while the DSv can be parked on the two ends of the vacuum chamber (100 mm of diameter and 250 mm of total length).}.

\subsection{Dynamic range of DSv}
\subsubsection*{Electromagnetic pick-up}

In principle, the DSv can measure signals from a single electron using an amplifier. However, since the ceramic PCB, on which the DS is mounted, is not shielded, when the beam passes by/through the PCB, a bipolar current pulse can be induced on the signal readout lines on the PCB. The high frequency signal given by such an induced current is called electromagnetic pick-up. 

\begin{figure*}[!tbh]
    \centering
    \includegraphics*[width=70mm]{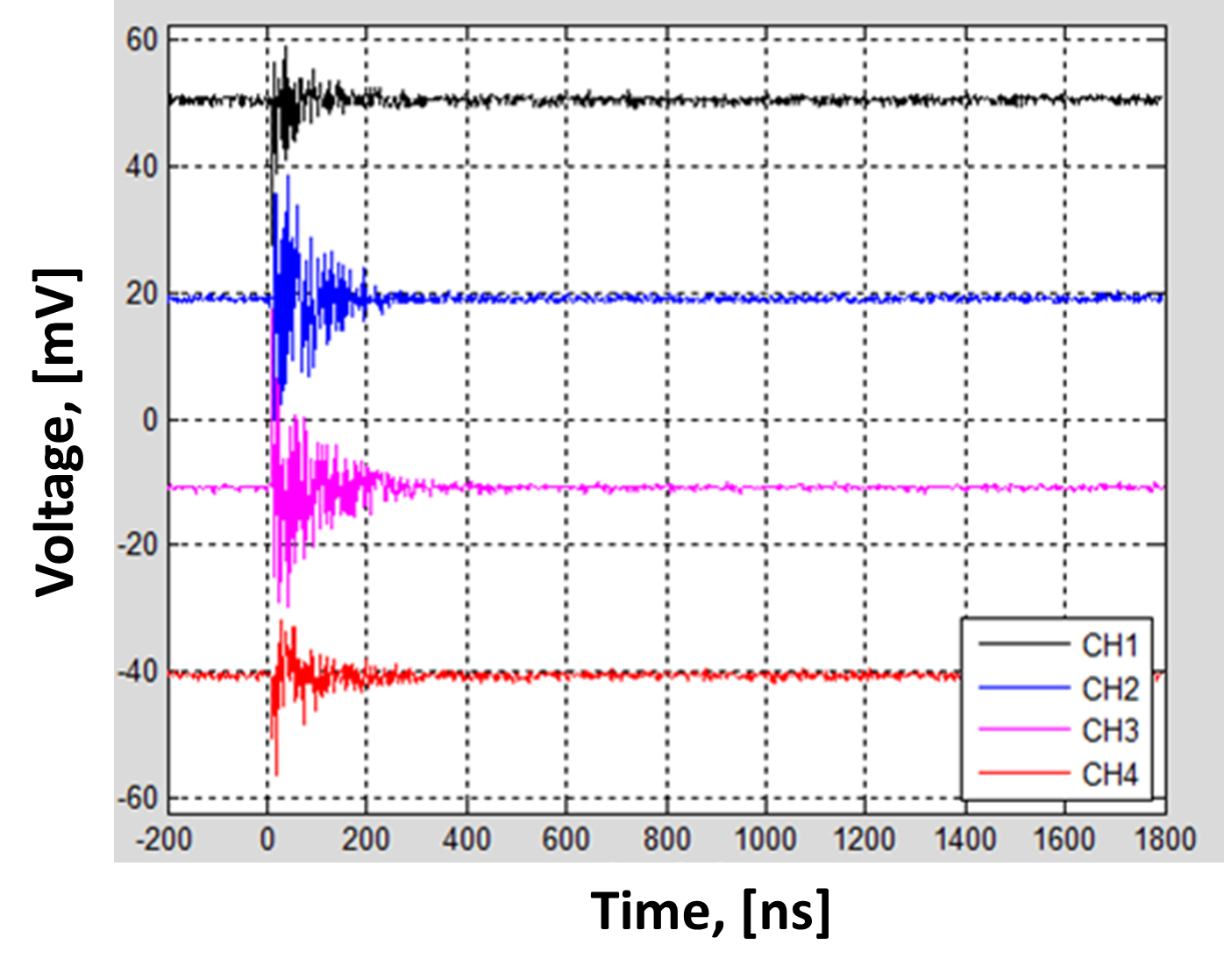}
     \caption{Pick-up signal on each channels taken on the 8\textsuperscript{th} December 2014.}
    \label{fig:pickup}
\end{figure*}

Pick-up signal was observed during the tests using the DSv without applying a bias voltage. An example of pick-up signal waveforms is shown in Fig.~\ref{fig:pickup}. The integrated charge is not absolutely zero but $\sim 3$ pC which correspond to $\sim 10^3$ electrons. This defines the lower limit of beam halo measurements for our present setup. This effect can be mitigated by shielding the PCB, which is one of the planned upgrades for the system. 

\subsubsection*{Linearity of DSv response}

The upper limit of DS performance has not been well defined by the experiment performed on several facilities. As mentioned before, linear responses up to $10^7$ and $10^9$ electrons/pulse have been demonstrated for the in air sCVD DS at PHIL with a 50~$\Omega$ readout system(see Fig.~\ref{fig:charge}) and for the in air pCVD DS at BTF with a 1~$\Omega$ readout system~\cite{steinresponse}, respectively. Besides, an \textit{in vacuum} beam halo monitor using pCVD diamond detectors at the SPring-8 Angstrom compact free-electron laser (SACLA) facility has also demonstrated a linear response up to $10^7$electrons/pulse with a 50~$\Omega$ readout system~\cite{aoyagi2012pulse}. However,``in vacuum" tests of diamond detector beyond $10^7$ electrons have not yet been described in any reference. Therefore, measurements were done at ATF2 with a total beam intensity of more than $10^9$ electrons to test the response of DSv. 

At ATF2, the input beam intensity for the DSv can be changed in two ways: 1) change the input laser intensity on the photo-cathode; 2) change the DSv position in the beam core. Integrated Current Transformers (ICT) are used at ATF2 to measure beams with more than $10^9$ electrons. To obtain lower beam intensities, we changed both the input laser intensity and the DSv position in the beam core. 

\begin{figure}[!htb]
   \centering
      \begin{subfigure}[b]{0.5\textwidth}
   \includegraphics*[width=65mm]{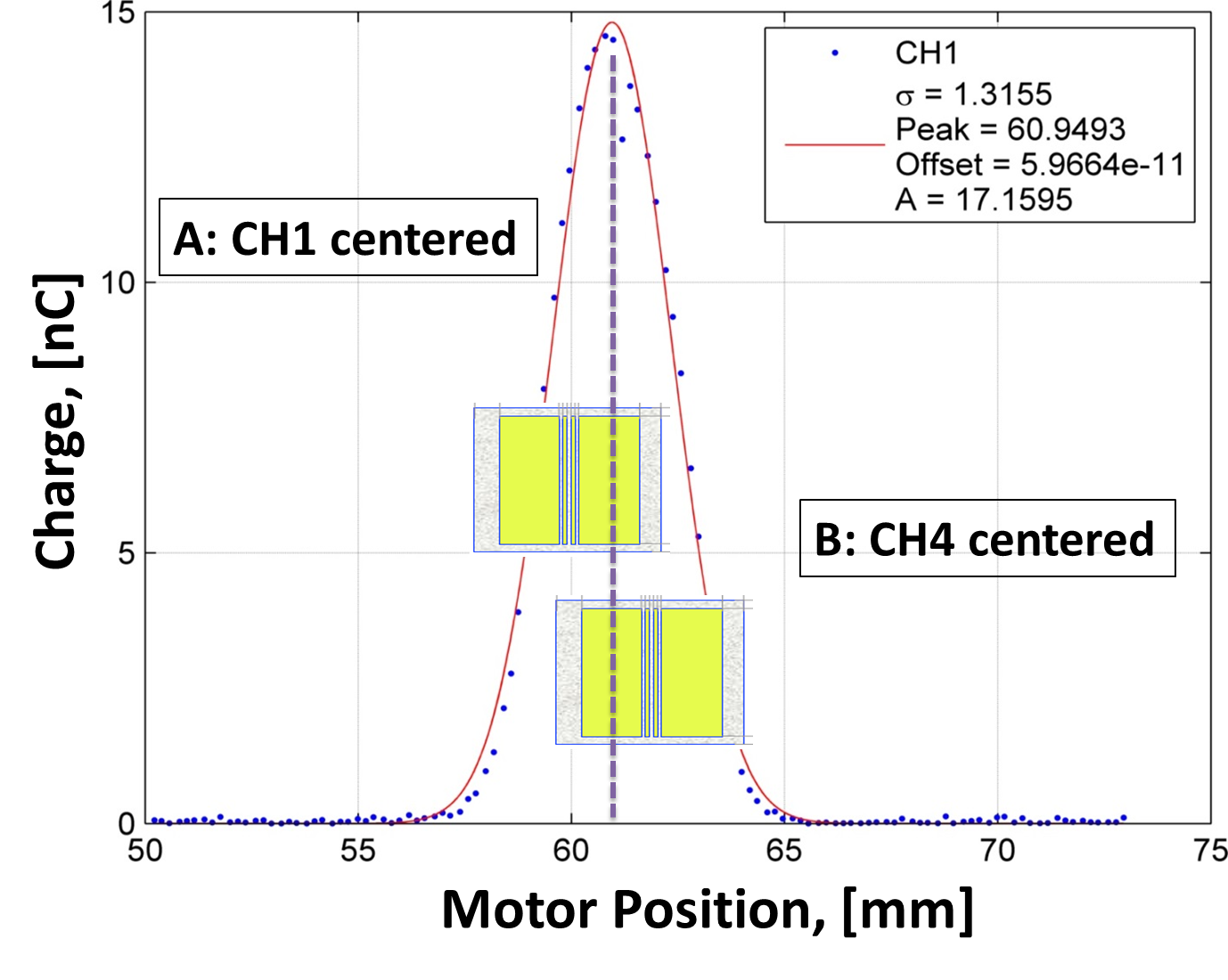}
                \caption{}  
       \end{subfigure}
   \begin{subfigure}[b]{0.5\textwidth}
   \includegraphics*[width=65mm]{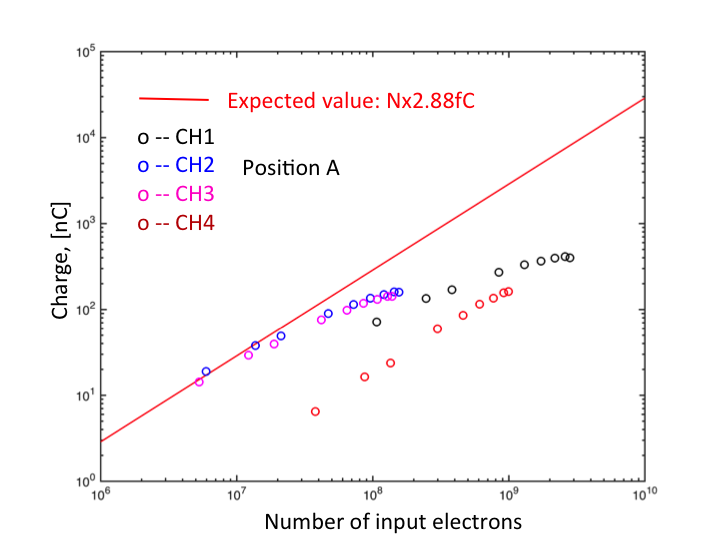}
                \caption{}  
       \end{subfigure}%
      \begin{subfigure}[b]{0.5\textwidth}
         \includegraphics*[width=65mm]{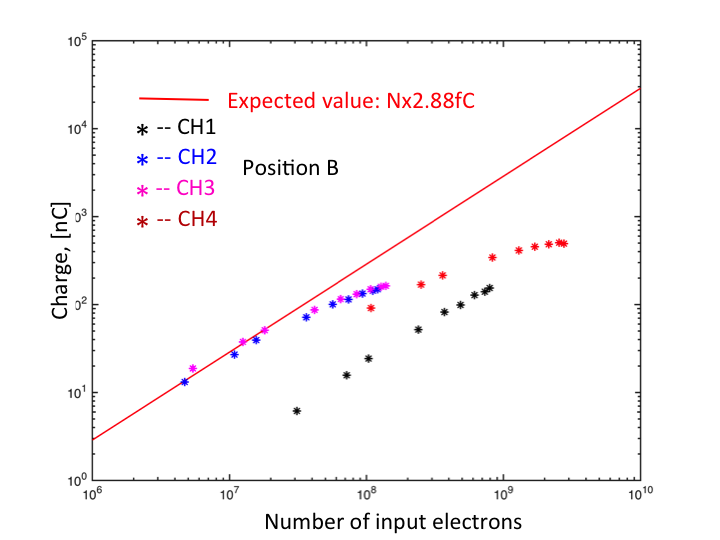}
                  \caption{}  
    \end{subfigure}
   \caption{Positions chosen for the measurements of the linearity of the response of the DSv (a) and the measurement results at position A (b) and B (c) for the horizontal DSv scanner.}
   \label{fig:CCE_HV}
\end{figure}

Fig.~\ref{fig:CCE_HV} (a) shows the positions chosen for the horizontal DSv scanner during the measurements. The DSv was moved to 2 positions (A and B) 2 mm apart from each other. The beam intensity was varied in the range from $10^9$ to $7\times10^9$ electrons and 100 waveforms were taken at each beam intensity. Before the measurements, the beam was aligned to the DSv center vertically, and then the beam core was scanned horizontally using the DSv and fitted by the convolution function of a Gaussian and a uniform distribution given by the strip width (see Eq.~\ref{S_result}). One example of the fit for the measured horizontal beam core distribution by CH1 is shown in Fig.\ref{fig:CCE_HV} (a).

For the calculation of the number of incident electrons, as the beam core follows a Gaussian distribution, the number of electrons intercepted on each channel $N_{CH}$ at different positions can be calculated as:
\begin{equation}
N_{CH}=A\int_{-\frac{1}{2}l_x+D_x}^{\frac{1}{2}l_x+D_x} \! e^{-\frac{x^2}{2\sigma _x^2} }\, \mathrm{d}x\int_{-\frac{1}{2}l_y}^{\frac{1}{2}l_y} \! e^{-\frac{y^2}{2\sigma _y^2} }\, \mathrm{d}y
\label{equation:Nch}
\end{equation}
where $A= N_{t}/(2\pi\cdot\sigma _x\cdot\sigma _y)$, $N_t$ is the total number of electrons in the beam, $l_x$ and $l_y$ are the horizontal and vertical widths of the strip and $D_x$ is the horizontal position offset from the beam center. The horizontal beam size ($\sigma _x$) in the formula is the measured beam size by the DSv, which can be obtained from the fit of the beam core. In the first measurements with the horizontal DSv, the vertical beam size ($\sigma _y$) could not be measured at the DSv location. It was then extrapolated from the post-IP wire scanner (post-IPW)\footnote{The post-IPW is located $\sim 60~cm$ downstream of the IP and upstream of the BDUMP bending magnet (see Fig.~\ref{fig:ATF2} and~\cite{WS_ATF2note}).} measured beam size.

Fig.~\ref{fig:CCE_HV} (b) and (c) shows the measured linearity response of the DSv at positions A and B. From the figure it can be seen that the charges collected by CH2 and CH3 are quite consistent, however, the CH1 and CH4 channels behave differently in the beam center and away from the beam center. At position A, where CH1 is centered in the beam, CH4 has less charge collected. The same behaviour can be observed for CH1 when CH4 is in the beam center at position B. The reason for this behaviour is under investigation.  

From both Fig.~\ref{fig:CCE_HV} (b) and (c)  it can be seen that the deviation of the signal from the linearity starts at a beam intensity of 10$^7$. This is consistent with previously obtained results using an in air DS (see Fig.~\ref{fig:charge}). It can be explained by the voltage drop on the 50~$\Omega$ resistor on the scope and the space-charge effect within the diamond bulk. 

Summarizing, the present DSv setup has a lower measurement limit of $\sim10^3$ electrons due to the electromagnetic pick-up and an operational range up to $\sim10^9$ electrons with a linear response up to $\sim10^7$ electrons. In order to estimate the effect of signal saturation on the beam size measurement, which is required for the normalization of the beam halo distribution, beam core scan were performed using different channel of the DSv.

\subsection{Beam core measurements}
Beam core measurements were performed at ATF2 for different optics. The ATF2 beam line is normally operated with a 10 times larger horizontal IP $\beta$-function optics than originally designed to reduce the effect of multipole field errors to a level comparable to the tolerances of the ILC final focus design~\cite{okugi2014linear}. This optics is referred to as the ``BX10BY1" optics because of the 10 times larger $\beta _x^*$ and the unchanged $\beta _y^*$ compared with the original design, while the original optics is labeled as ``BX1BY1" optics. Small (``BX10BY0.5") and large (``BX100BY1000") $\beta$-function optics are also used at times for different studies. 

The beam core is scanned by the DSv with a 30 dB attenuator added to each channel and with a bias voltage of -400 V. For the horizontal and vertical beam core scans, the beam is aligned vertically and horizontally, respectively to the center of the corresponding DSv. 

The measured distribution ($S(\xi )$) can be considered as the convolution of a uniform distribution (\textit{L(x)}) and a Gaussian distribution (\textit{I(x)})~\cite{king2013analysis}, it gives:

\begin{equation}
S(\xi )=  K(L*I)(\xi )= KI_0\int_{-\xi-\frac{l}{2} }^{\xi +\frac{l}{2} } \! e^{ -\frac{(x-\mu)^2}{2\sigma ^2}} \, \mathrm{d}x = KI_0(erf(\frac{\xi +\frac{l}{2}-\mu}{\sqrt{2}\sigma})-erf(\frac{\xi -\frac{l}{2}-\mu}{\sqrt{2}\sigma}))
\label{S_result}
\end{equation}
where $K$ and $I_0$ are constants and $\ell$ is the width of the DSv strips. 

\begin{figure*}[htb]
    \centering
    \includegraphics*[width=80mm]{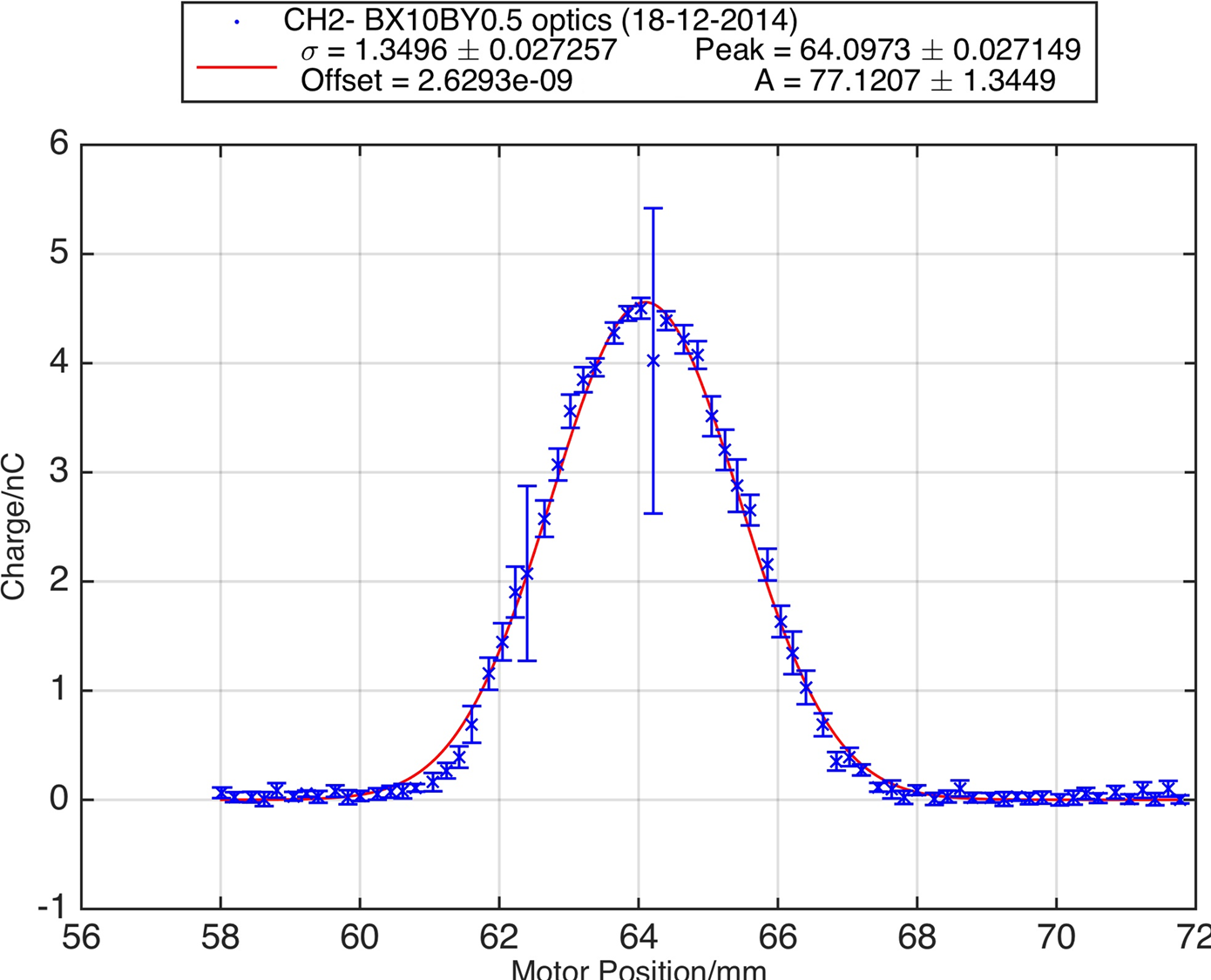}
     \caption{Example of horizontal beam core fit for CH2.}
    \label{fig:DSv_CH2}
\end{figure*}

The measured beam core signal is fitted using Eq.~\ref{S_result}. An example of horizontal beam core fit is shown in Fig.~\ref{fig:DSv_CH2}. The fitted beam size ($\sigma$) by this Eq.~\ref{S_result} is corrected by the strip width ($\mathit{l}$). 

The beam sizes measured by different channels are compared in Table \ref{table:beamsize} for different beam optics. From the table it can be seen that the differences in the beam sizes measured by different channels are within $13 \%$. This level of difference is acceptable for the beam halo normalization for the first studies of aperture effects and beam halo dependence on beam intensity and vacuum levels, which will be explained in more detail in the following sections. However, to enable the parametrization of the beam halo distribution, similar to what was done using the wire scanners at ATF2~\cite{suehara2008design,WS_ATF2note}, further study of the saturation effects and cross-talk between the channels may be required.

\begin{table}[hbt]
   \centering
   \begin{tabular}{|l|l|l|l|}
   \hline
      \textbf{} &{BX10BY0.5,[mm]} &{BX10BY1,[mm]} &{BX100BY1000,[mm]}  \\ \hline
CH1  & $1.401 \pm 0.047$ &$1.300 \pm 0.046$ &$1.261 \pm 0.026$  \\ \hline
CH2  & $1.350 \pm 0.027$ 	&$1.362 \pm 0.061$  &$1.254 \pm 0.025$  \\ \hline
CH3  & $1.261 \pm 0.028$	&$1.296 \pm 0.063$  &$1.208 \pm 0.030$  \\ \hline
CH4 & $1.351 \pm 0.023$ &$1.272 \pm 0.054$ & $1.120 \pm 0.031$ \\ \hline

  \end{tabular} 
    \caption{Comparison of measured beam sizes for different optics by different channels of the DSv}
    \label{table:beamsize}
\end{table}

In order to verify the DSv measured beam sizes, we compared them with the values obtained from the MAD-X simulation~\cite{schmidt2005mad} and with the measured beam sizes using the post-IPW wire scanner. Since the narrow strips are less affected by the saturation of the collected charge in the beam core, it is preferable to use their measured beam size for the comparison. Table. \ref{table:beam_parameter_3} shows the comparison between simulated and measured beam sizes for the post-IPW wire scanner (see Fig.~\ref{fig:ATF2}) and the DSv. It can be seen that the discrepancy between simulated and measured beam size is larger for the post-IPW than for the DSv. It could be that the DSv (CH2) provides a more reliable beam size measurement. 

\begin{table}[hbt]
   \centering
   \begin{tabular}{|l|l|l|l|l|l|l|}
   \hline
      \textbf{} &{BX10BY0.5,[mm]} &{BX10BY1,[mm]} &{BX100BY1000,[mm]} \\ \hline
post-IPW simulated  & 0.156 & 0.156 &8.744e-02 \\ \hline
post-IPW measured & 0.197&0.266 & 6.150e-02 \\ \hline
DSv simulated & 1.394 	 &1.395	&1.077\\ \hline
DSv measured(CH2)& $1.350\pm 0.027$	 &$1.362\pm 0.061$ &$1.254\pm 0.025$ \\ \hline

  \end{tabular} 
    \caption{Comparison of simulated and measured horizontal beam sizes at the post-IPW wire scanner and the DSv for different optics (the errors on post-IPW measured beam sizes are negligible).}
    \label{table:beam_parameter_3}
\end{table}


\subsection{Beam halo measurements}

First beam halo measurements were performed in both horizontal and vertical planes to check the effect of apertures along the beam line, as well as to study the horizontal beam halo dependence on beam intensity and the vertical beam halo dependence on the vacuum level in the ATF damping ring.

\begin{figure*}[!tbh]
    \centering
    \includegraphics*[width=75mm]{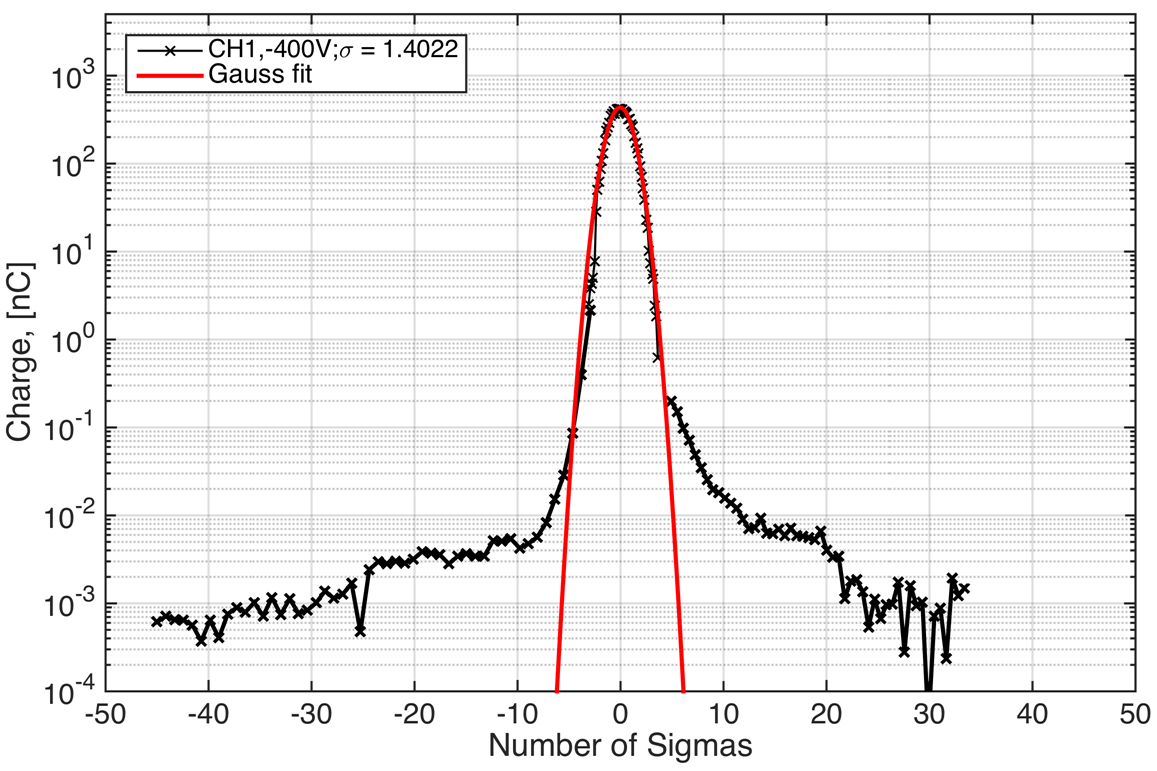}
    \includegraphics*[width=75mm]{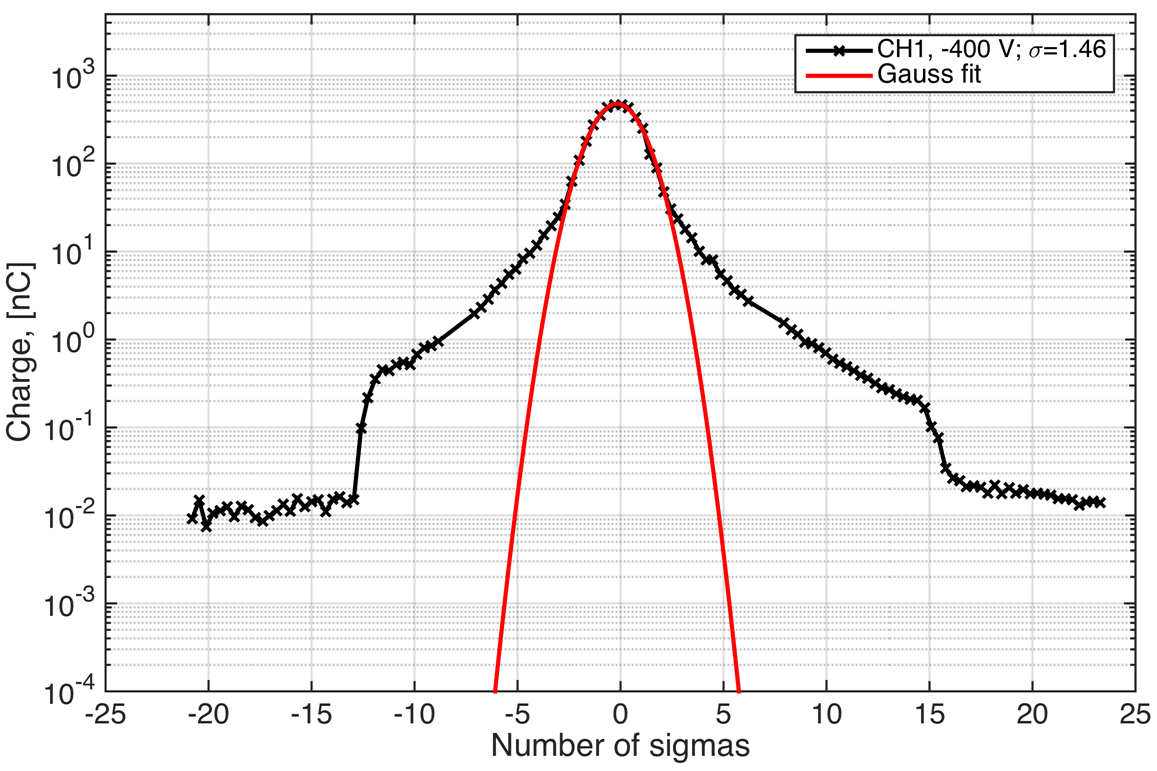}
     \caption{Measured beam core and beam halo distribution using DSv with cut by the apertures in the horizontal (left, BX10BY0.5 optics) and vertical (right, BX10BY1 optics) plane.}
    \label{fig:Cut_BDUMP}
\end{figure*}

An example of combined beam core and beam halo distribution for the data taken with CH1 is shown in Fig.\ref{fig:Cut_BDUMP} for horizontal (left) and vertical (right) scans, respectively. The measured beam size for the beam core is $\sim$1.4 mm and $\sim$1.46 mm in horizontal and vertical plane, respectively. The measured charge for the beam core is normalized by the factor corresponding to 30 dB attenuation. It can be seen that in the horizontal measurement, a dynamic range of $\sim 10^6$ has been successfully achieved. The noise level in this case is at the level of $\sim 10^{-3}$ nC, which is mainly caused by the electromagnetic pick-up as mentioned before. 

\subsubsection*{Cut of beam halo by apertures}

In Fig.\ref{fig:Cut_BDUMP}  rather sharp edges can be seen on both sides of the beam halo distribution. Although there is no dedicated beam halo collimator in the ATF2 beam line at present, the apertures of kickers, magnets, BPMs and beam pipes on the beam line can intercept the beam halo. In order to study and define the possible beam halo cut positions by the present apertures in the ATF2 beam line, simulations using MAD-X have been performed. The simulated cut values for several locations with tight apertures and large $\beta_{x,y}$ are shown in Table \ref{table:Cuth} and Table~\ref{table:Cutv} for horizontal and vertical beam halo, respectively. The simulated and measured cut values at the DSv location are also shown in the Table. 

\begin{table}[h!]
   \centering
   \begin{tabular}{|l|l|l|l|}
   \hline
  
      \textbf{} &\multicolumn{3}{c|}{Horizontal Cut, [$\sigma_x$]} \\ \hline
            \textbf{} &\multicolumn{2}{c|}{Simulated cut by apertures} &\multicolumn{1}{c|}{Measured cut}\\ \hline
      &   MSF5FF  & DSv & DSv  \\
       &  (72.23 m) & (93.51 m) &(93.51 m)  \\  \hline
      
BX10BY0.5         & $\pm16.01$ &$\pm25.86$ & - \\ \hline
BX10BY1  &  $\pm15.99$ & $\pm25.84$  & -24/+20 \\ \hline

  \end{tabular} 
    \caption{Simulated horizontal cut of beam halo by apertures at MSF5FF (C-Band BPMs for sextupole SF5 in the final focus line) and the DSv location (in number of $\sigma _{x}$) and measured cut at DSv location (given separately for the left (low energy) and right (high energy) side, see Fig.~\ref{fig:Cut_BDUMP}). Their distance from the entrance of the beam line is shown in the table.    
    }
    \label{table:Cuth}
\end{table}

From Table~\ref{table:Cuth} we can see that the tightest horizontal cut is at the MSF5FF location. This cut is given by the C-Band BPMs (with radius of 8 mm) installed next to the sextupole SF5 at 72.2 m from the beginning of the beam line (see Fig.~\ref{fig:ATF2}). They intercept the beam halo at 16.0$\sigma_x$. And at the DSv location, which is 1.35 m from the exit of the BDUMP magnet, the cut is around 25.9$\sigma_x$. This cut is caused by the beam pipe installed after the BDUMP with a radius of 31.5 mm. Since the cut of beam halo by the C-Band BPMs is tighter than that from the beam pipe after BDUMP, we would expect to see this cut on the horizontal beam halo distribution at the DSv location. 

However, in the first measurements, only the cut given by the beam pipe installed after the BDUMP exit at -24$\sigma_x$ and +20$\sigma_x$ ($\sigma_x$ measured by CH1) is visible (see Fig. \ref{fig:Cut_BDUMP} (left))\footnote{This cut was verified experimentally by displace the beam horizontally using the BDUMP bending magnet.}. Further measurements should be performed upstream of the IP to check the effect of the apertures of the C-Band BPMs.

In the vertical plane (see Table~\ref{table:Cutv}), a tappered beam pipe (TBP) was installed 61.4 m from the beginning of the beam line (see Fig.~\ref{fig:ATF2}), between the QD10B and QD10A magnets, where the vertical beam size is relatively large. The radius of this TBP is 8 mm. It can serve to reduce the background for the IPBSM without exciting wakefields. The vertical cut by the TBP is estimated to be at $\sim$20$\sigma_y$ for BX10BY0.5 optics and $\sim$27$\sigma_y$ for BX10BY1 optics.  As the beam is focused to a few tens of nm at the IP, the beam divergence is very large after the IP. Thus the tightest vertical cut is given by the BDUMP bending magnet. At the exit of this magnet, the cut goes down to $\sim$10$\sigma_y$ and $\sim$14$\sigma_y$, for the BX10BY0.5 and BX10BY1 optics, respectively. 

\begin{table}[h!]
   \centering        
   \begin{tabular}{|l|l|l|l|l|}
   \hline
  
      \textbf{} &\multicolumn{4}{c|}{Vertical Cut, [$\sigma_y$]} \\ \hline
            \textbf{} &\multicolumn{3}{c|}{Simulated cut by apertures}&\multicolumn{1}{c|}{Measured cut} \\ \hline
      & TBP & BDUMP & DSv  & DSv \\
        & (61.39 m) & (92.16 m) & (93.51 m) & (93.51 m) \\  \hline
      
BX10BY0.5         & $\pm20.06$  &$\pm10.15$ & $\pm20.17$ &- \\ \hline
BX10BY1 & $\pm26.85$   & $\pm13.59$ & $\pm26.994$ & -12/+15   \\ \hline

  \end{tabular} 
    \caption{Simulated vertical cut of beam halo by apertures at TBP (Tapered Beam Pipe), BDUMP and DSv location (in number of $\sigma _{y}$) and measured cut at DSv location (given separately for the upper and lower side, see Fig.~\ref{fig:Cut_BDUMP}). Their distance from the entrance of the beam line is shown in the table.}
    \label{table:Cutv}
\end{table}

The measured vertical beam core and beam halo distri\- butions for the BX10BY1 optics are shown in Fig. \ref{fig:Cut_BDUMP} (right) with cut positions at -12$\sigma_y$ and +15$\sigma_y$\footnote{This cut was verified experimentally by displace the beam vertically using the vertical shift on QD0FF.}, which is quite consistent with the predicted cut value by the BDUMP bending magnet ($13.6\sigma_y$ for the BX10BY1 optics). This cut on beam halo may generate large background for beam halo measurements at the DSv location \cite{Illia_ATF2note}. The TBP may be moved vertically to limit such background. However, a dedicated beam halo collimation system is being prepared upstream for this purpose \cite{Nuria_IPAC2015}. 

\subsubsection*{Horizontal beam halo dependence on beam intensity}
\begin{figure*}[!tbh]
    \centering
    \includegraphics*[width=75mm]{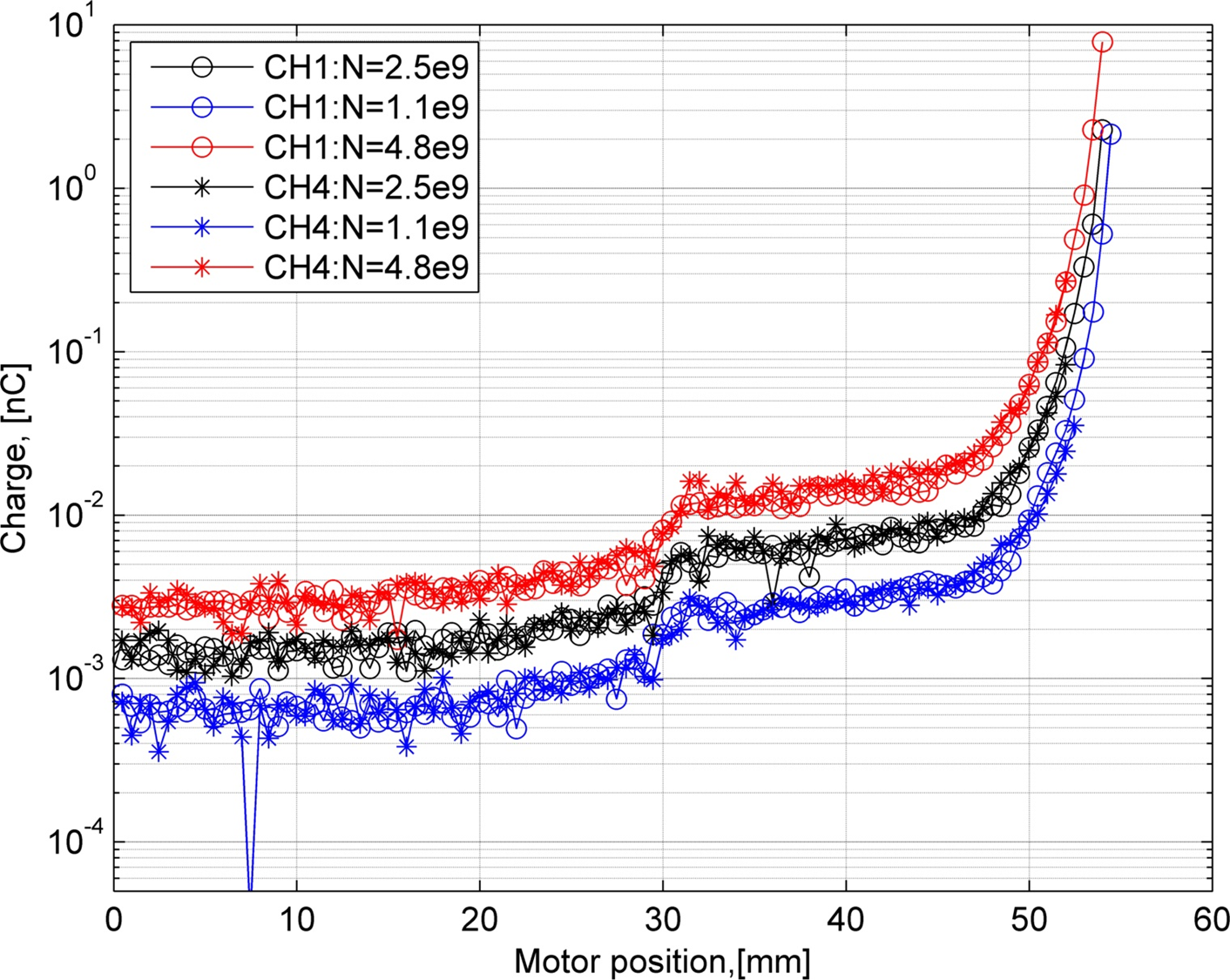}
    \includegraphics*[width=75mm]    
    {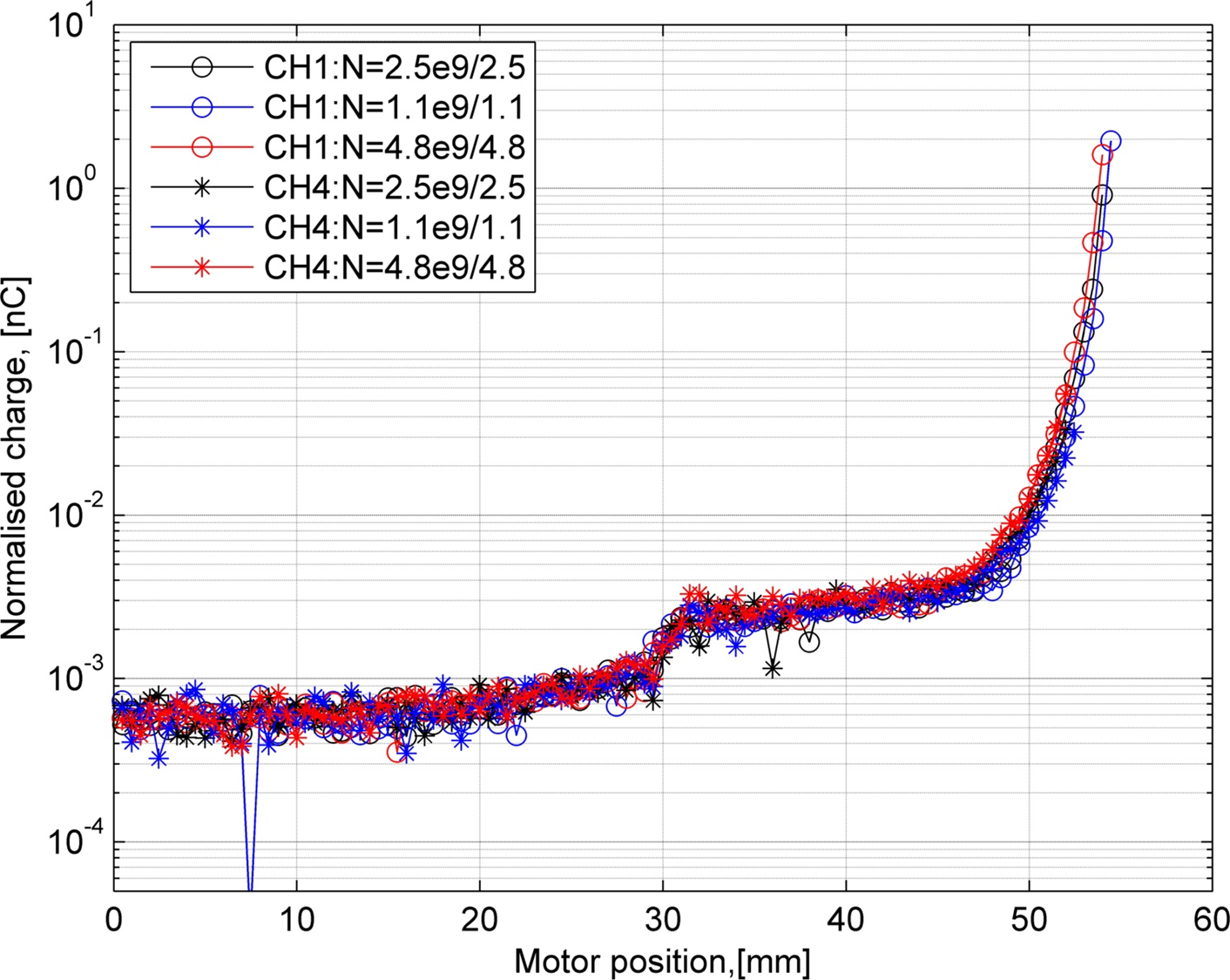}
     \caption{Horizontal beam halo distribution measured for different beam intensity before (left) and after (right) normalization on the low energy side.}
      \label{fig:Intensity}
\end{figure*}

The effect of beam intensity changes on beam halo distributions was studied by changing the total beam intensity. The total beam intensity can be controlled by changing the intensity of the laser which is injected into the photon gun. In principle the laser intensity can be changed from 0$\%$ to 100$\%$ and the beam intensity can be changed from 0 to 10$^{10}$, however, as the ICT, which was used for the beam intensity measurement, is not sensitive to a beam intensity below 10$^9$ e$^-$, in this measurement we changed the beam intensity only from 1.1$\times$10$^9$ to 2.5$\times$10$^9$ and to 4.5$\times$10$^9$ e$^-$. The measurement results are shown in Fig. \ref{fig:Intensity}.

An increase of beam halo population can be observed after increasing the beam intensity in Fig.~\ref{fig:Intensity} (left). However, from Fig.~\ref{fig:Intensity} (right), it can be seen that after normalising the beam intensity to $10^9$ for all the data, the shapes of the beam halo distributions are quite consistent for different intensities. Thus, we can conclude that for the intensity changes that we have applied, there is negligible change on horizontal beam halo distributions.
\subsubsection*{Vertical beam halo dependence on vacuum level}

\begin{figure*}[!tbh]
    \centering
    \includegraphics*[width=80mm]{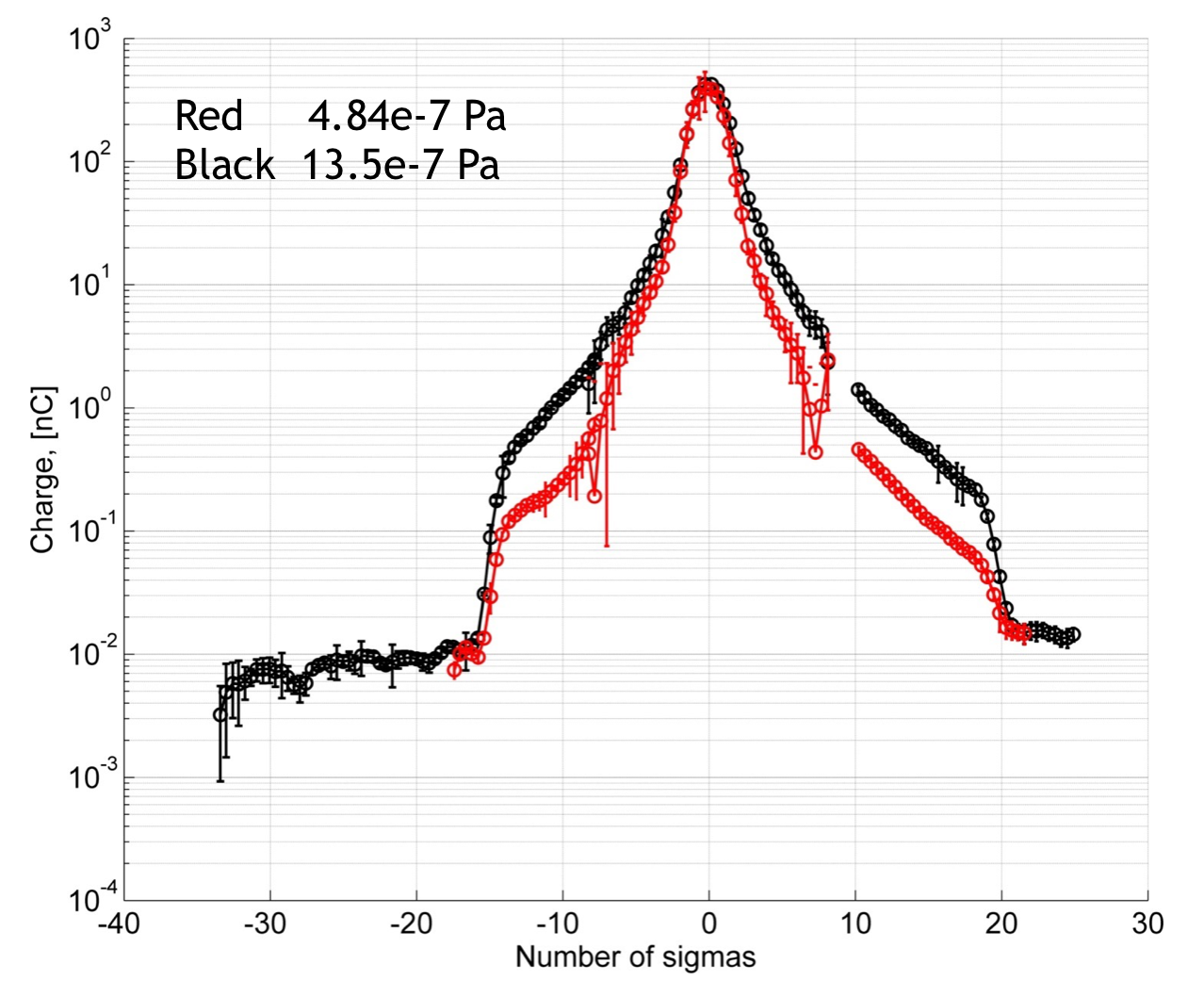}
     \caption{Vertical beam core and beam halo distribution measured for two different vacuum levels: 4.84$\times10^{-7}$ Pa (red) and 13.5$\times10^{-7}$ Pa (black).}
      \label{fig:Vacuum}
\end{figure*}

Beam halo can be generated by different processes in the ATF damping ring (DR), such as beam-gas scattering, beam-gas bremsstrahlung and intra-beam scattering. It was pointed out in Ref.~\cite{Dou2014analytical} that the transverse halo generation is dominated by beam gas scattering. This can be verified experimentally by changing the vacuum level in the DR. Fig. \ref{fig:Vacuum} shows an example of vertical beam halo measurements, done with part of the ion pumps in the DR turned off to obtain a relatively high vacuum level of $1.35\times 10^{-6}$ Pa. A second measurement was done subsequently after turning the ion pumps back on to recover the nominal value of $4.84\times 10^{-7}$ Pa. 

It can be seen from this first measurements, that the beam halo population increases more or less proportionally with the vacuum level. Similar measurement results were obtained using the YAG:CE screen installed in the extraction line of ATF2 as described in Ref.~\cite{naito2015beam}. Both measurement results have shown a consistent beam halo dependance on vacuum level with the analytical results obtained in Ref.~\cite{Dou2014analytical}.

\section{Conclusion and Prospects}
An \textit{in vacuum} diamond sensor (DSv) with large dynamic range was successfully developed and implemented for beam halo measurements at ATF2.  A dynamic range of $\sim10^6$ was obtained, allowing simultaneous measurement of beam core and beam halo. The DSv has been tested with very high beam intensities ($10^9~e^-$). Saturation effects of the present DSv setup have been observed and investigated. These studies are very valuable for the development of diamond detectors operating with very high intercepted charge.

For the characterization of the in air diamond sensor sample, tests using an $\alpha$ source and electron beam were performed. Charge collection efficiency, mobility, saturation velocity and lifetime of the diamond sensor have been studied using the Transient Current Technique (TCT). The saturation velocity was compared with the linear extrapolation from the signal amplitude measured at PHIL, it showed a consistent result with the experimental observation at PHIL. The decrease of charge collection efficiency (CCE) observed at very high beam intensity ($>10^7~e^-$) due to voltage drop and the limited lifetime of charge carriers was explained. 

Characterizations of DSv were carried out using the electron beam at ATF2, where a linear response up to $10^7$ electrons was confirmed, in consistancy with the measurements done at PHIL using the in air DS. In the present design, the performance of the DSv is limited by the electromagnetic pick-up at the level of $10^3$ electrons and by non-linearity starting from $10^7$ electrons. For the signal pick-up, shielding of the PCB can be applied. The non-linear response of the DSv due to large voltage drops in the 50~$\Omega$ resistor in the readout channel may be avoided by adding a smaller resistor in parallel. With these solutions, the dynamic range of the DSv can be further improved. 

Performance of the DSv for beam core measurements was confirmed by comparing the measured beam sizes with similar measurements using the post-IP wire scanner. First horizontal beam halo measurements were performed for different beam intensities, confirming the expected proportional increase in beam halo. In the vertical plane, beam halo measurements were performed for different vacuum levels. An increase of beam halo population with the vacuum level was observed as expected from the analytical estimation~\cite{Dou2014analytical}.

In the future, beam halo measurements using the DSv will provide important quantitative information on the dependence of the beam halo distribution on beam intensities for different beam optics, in particular for the ultra low $\beta _y$ optics study, which is important for both ILC and CLIC, where high luminosity ($\sim10^{34}~cm^{-2}\cdot s^{-1}$) will be obtained with ultra low $\beta _y$. Such ultra low $\beta _y$ will be tested at ATF2 and the background for these tests needs to be mastered. Moreover, the DSv measured beam halo distributions will be compared both with the measurements using other instruments~\cite{Nuria_IPAC2015,naito2015beam} and with theoretical models~\cite{Dou2014analytical} to distinguish different beam halo generating mechanism and to optimise the collimation system.

\section{Acknowledgments} 

This work is supported by China Scholarship Council (CSC), P2IO LABEX Physique des 2 Infinis et des Origines and the associated international laboratories: France China Particle Physics Laboratory (FCPPL) and Toshiko Yuasa France Japan Particle Physics Laboratory (TYL-FJPPL). The authors would like to thank all the members of the ATF2 collaboration for their help during the experiments, especially for the KEK staff, T. Okugi, T. Naito, K. Kubo and S. Kuroda. We also would like to express our acknowledgement to all the PHIL operators and the vacuum group at LAL, especially H. Monard, P. Lepercq, S. Chancé, J. N. Cayla, T. Vinatier and B. Mercier, for their help during the tests of the diamond sensor and the mechanical system at PHIL.

\newpage


 \bibliographystyle{elsarticle-num} 
 \bibliography{references}





\end{document}